\documentstyle{Iaa}

\def\tvi(#1,#2){\vrule height #1pt depth #2pt width 0pt}
\def\p{\partial}
\def\od{\omega_{\Delta}}
\def\tx{\tilde\xi}
\def\bk{\bar k}
\def\bx{\bar\xi}
\def\cx{\check\xi}
\def\tom{\tilde\omega}
\def\tod{\tilde\omega_{\Delta}}
\def\todp{\tilde\omega_{\delta}}

\def\e{{\rm e}}
\def\d{{\rm d}}
\def\ie{i.e. }
\def\eg{e.g. }
\def\etal{et al. }

\newlength{\largeur}
\newlength{\saut}
\def\marge#1{ \setlength{\largeur}{\columnwidth}
 \addtolength{\largeur}{-#1}
 \setlength{\saut}{0.5\largeur}\hspace*{\saut}}
\def\picture #1 by #2 (#3){
 \marge{#1} \vbox to #2{
  \hrule width #1 height 0pt depth 0pt
  \vfill
  \special{picture #3}}}
\def\scaledpicture #1 by #2 (#3 scaled #4){{
  \dimen0=#1 \dimen1=#2
  \divide\dimen0 by 1000 \multiply\dimen0 by #4
  \divide\dimen1 by 1000 \multiply\dimen1 by #4
  \picture \dimen0 by \dimen1 (#3 scaled #4)}}
\begin{document}

\thesaurus{12( 02.01.2;02.09.1;02.13.2;09.03.2;11.09.4;11.13.2)}
\title{The Parker instability in disks with differential rotation}

\author{T.~Foglizzo \and M.~Tagger}

\offprints{T.~Foglizzo}

\institute {Service d'Astrophysique, CE-Saclay, 91191 Gif-sur-Yvette
Cedex, France}

\date{Received 1993 December 21; accepted 1994 February 23}
\maketitle
%---------------------------------------

\begin{abstract}
We present a detailed study of the growth of the Parker instability in a
differentially rotating disk embedded in an azimuthal equilibrium
magnetic field, such as the interstellar gas or an accretion disk.
Basic properties of the instability without shear are first recalled.
Differential rotation is modeled in the shearing sheet approximation,
classical in the theory of spiral density waves,  and the linearized
problem is transformed into a second order differential equation. The
action of differential rotation is reduced to two different effects, (i)
a linear time-dependence of the radial wavenumber, and (ii) a  radial
differential force. We present both exact numerical solutions, and
approximate analytical ones based on the WKB approximation in the limit
of weak differential rotation. This allows us to discuss the
mathematical and physical nature of new behaviours obtained numerically
in cases with realistic differential rotation. Most important are (i) a
transient  natural stabilization of the Parker mode due to the radial
differential force (ii) the generation of magnetosonic waves (\ie spiral
density waves if we had included self-gravity) and Alfvenic ones, and
(iii) in a certain parameter range a possible ``turn-over'' of the
perturbation whereby, quite surprisingly, matter which had started being
elevated by the instability may end up dropping towards the disk
midplane. A simplified model shows the possible observable effects of
this turn-over.


\end{abstract}
%---------------------------------------

\section{Introduction}
The Parker instability has often been invoked in the formation of
molecular clouds along the spiral arms of our galaxy (Blitz \& Shu 1980;
Gomez de Castro \& Pudritz 1992), as well as in external galaxies (Rand
\& Kulkarni 1990); it is also considered as a prime agent in the
turbulent dynamo theory for the generation of the galactic magnetic
field (Ruzmaikin \etal 1988); this theory, which relies on magnetic
loops rising  from the galactic disk by the Parker instability (Parker
1992) and being twisted by differential rotation to finally be cut off
by magnetic reconnection, is gaining increasing observational evidence
as  such loops are observed in the halo above the galactic plane of
nearby spiral galaxies (Beck 1991).\\
In accretion disks, if they are sufficiently ionized, one may also expect
to have a strong magnetic field, both primordial (the magnetic flux from
the interstellar field frozen in the accreting matter) and coming from a
turbulent dynamo. Furthermore, recent works (Tagger \etal 1990, 1992;
Balbus and Hawley 1991) have shown that in presence of this field such
disks can become extremely unstable, providing a long-awaited hint to
the mechanism of turbulent accretion. However the basic magnetic
configuration that we will consider here, that of a purely toroidal (\ie
azimuthal), stratified field, has not yet been studied in this
context.\\
To our knowledge, the Parker instability has been studied only in the
presence of solid (Zweibel and Kulsrud 1975) or weakly differential (Shu
1974) rotation, mainly establishing that no overall stabilization could
be expected.  In particular Shu (1974) showed that differential rotation
resulted in a  coupling of the normal mode solutions, and restricted his
study to the ``weak-shear limit'' to allow a normal mode analysis~: he
derived dispersion relations in the two cases of (i) uniform rotation
and (ii) differential rotation at infinite $k_x$, and showed the latter
case to be independent of the intensity of shear. The analytical
difficulties introduced by the  modelization of differential rotation
are responsible for the lack of a more detailed description.\\
Differential rotation has two main effects, which will guide our
discussion in this paper: the first one is that, in the same manner as
spiral density waves, the perturbation is sheared by the fluid motions,
so that its radial wavenumber evolves with time. It is however important
to note here an important difference with spiral density waves: the
latter travel radially in the disk and are amplified when they reach
small radial wavenumbers; the Parker instability, on the other hand, is
stationary and is known to be more strongly amplified at large radial
wavenumbers; thus, even though the shearing motion can take it to small
radial wavenumbers where it is less unstable (or even stable, as a
consequence of the second effect of differential rotation), this can be
only a transient effect and ultimately it gets back to large wavenumbers
and strong amplification.\\
The shearing motion will be shown to naturally introduce a linear
coupling between the modes, by which the evolution of the unstable
Parker mode generates two  waves : a magnetosonic one (\ie a spiral
density wave), traveling radially, and an alfvenic one traveling along
the field. \\
The second effect of differential rotation appears in the polarization
of the wave, \ie the direction of the perturbed motions~: without
rotation, the motion would be essentially vertical (the rising of the
flux tubes) and azimuthal (the motion of the gas along the field lines),
with a small radial component (the expansion of the flux tubes). With
differential rotation the azimuthal motion creates centrifugal and
Coriolis forces, leading to radial motions which twist the flux tubes.
The complex interplay of these forces (and of the restoring force from
magnetic tension) explains the new behaviors we will find, \eg a
transient stabilization of the Parker instability and, if shear is high
enough, a ``turnover'' of the waves, whereby the matter that had started
to be elevated by the instability may end up dropping towards the disk
midplane.\\
We first describe in Sect.~2 the general hypotheses of our modelization:
the classical ones used to describe the Parker instability in a
stratified equilibrium, and the shearing sheet model used to describe
waves in a differentially rotating disk; we summarize in Sect.~3 the
classical features of the Parker instability. We introduce in Sect.~4 the
cosmic-ray pressure modelized in the same way as Shu (1974), and recall
in Sect.~5 the effect of the Coriolis force on the Parker instability in
a disk with uniform rotation.\\
Differential rotation is introduced in Sect.~6. The differential system
obtained is solved numerically in Sect.~7 to illustrate the possible
transient stabilizing influence of shear, the linear coupling to
oscillatory modes, and the turn-over. \\
Section~8 contains our main analytical calculations, allowed by an
appropriate mathematical transformation (Sect.~8.1).  We give in
Sect.~8.2 an original analytic proof of the fact that in the
configuration we are studying, that of a purely toroidal field, the
behaviours are always asymptotically dominated by the Parker
instability. This analytical treatment permits a description of the
asymptotic effect of differential rotation in terms of coupling
constants between the three asymptotic modes.\\
A WKB approximation in the case of low shear allows us to get into the
details of the transient shearing stabilization in Sect.~8.3. The linear
coupling to oscillatory modes is approached in Sect.~8.4, and the
possible turn-over is formulated analytically in Sect.~8.5.\\
Section~9 deals with the evolution of the most likely type of initial
perturbation, \ie one that is radially localized. We present a
modelization allowing us to describe its behavior in time, and the
consequences of the turn-over mechanism.\\
A parallel with the shearing instability in accretion disks with a
vertical component of the magnetic field is drawn in Sect.~10, outlining
the ambivalent nature of differential rotation according to the strength
of the magnetic field. We show that the transient effect of a strong
differential rotation can lead to an over-amplification of Parker
instability, although a weak differential rotation is  stabilizing.\\
A final discussion and conclusions are given in Sect.~11.\\
In order to facilitate the discussion of the physical mechanism and the
main results, all complex calculations are given in separate appendices.
Furthermore, since very little is known about the magnetic field in
accretion disks, we will, throughout the paper, refer to the
configuration and observed properties of galactic disks.

\section{General hypotheses}

\subsection{Hydrostatic equilibrium}
We adopt the same framework as most authors as regards the  equilibrium
equations. The self-gravity of the gas is  neglected, though it is
supposed to play an important role at scales  much smaller than the
scales considered here, which are of the order of the scale height of
the disk (Elmegreen 1982a,b, 1987, 1989, 1991; Hanawa \etal 1992).\\
The vector ${\bf x}$ is along the radial direction, ${\bf y}$ is along
the azimuthal magnetic field (${\bf B}_o=B_o {\bf y}$) and ${\bf z}$ is
along the axis of galactic rotation (${\bf \Omega}=\Omega{\bf z}$).\\
The vertical component of the gravitational field is approximated, for
the sake of simplicity, as a constant $ -{\rm sign}(z)g_z {\bf z}$,
although it is known to  grow linearly in the first 100 pc of the
galactic disk. More  realistic vertical profiles do not bring any
significant qualitative change (Horiuchi \etal 1988; Giz \& Shu
1993).\\
The magnetic and cosmic ray pressures are assumed to be proportional to
the gas thermal pressure, with the usual constants $\alpha$ and
$\beta$~:\\
\begin{equation}
P_{b}=\alpha P_{\rm th} \mbox{ and } P_{\rm cr}=\beta
P_{\rm th}.
\end{equation}
This is not very realistic according to recent observations~:  the scale
height of the magnetic field and hence that of the cosmic ray gas is now
known to be much larger than that of the HI gas (Bloemen 1987; Boulares
\& Cox 1990). Our simplification may lead to an over-estimation, in the
lower regions of the galactic disk, of the gradient of the magnetic field
strength responsible for the Parker instability. This discrepancy in
scaleheights however may be seen as the consequence of the instability
which tends to blow the magnetic field lines towards the halo (Zweibel
\& Kulsrud 1975; Parker 1975).\\
The adiabatic index of the galactic gas is taken to be $\gamma=1$ so as
to avoid the interchange instability (Parker 1966). The thermal pressure
is simply~: $P_{\rm th}=\rho_o(z) a^2$ , where $a$ is the sound speed in
the gas, assumed to be independent of height.\\
The hydrostatic vertical equilibrium is given by~:
\begin{equation}
{d\over dz}\left(P_{\rm th}+P_{b}+P_{\rm
cr}\right)=-\rho_o(z) g_z(z),
\end{equation}
which leads to~:
\begin{equation}
\rho_o(z) = \rho_o(0) \exp(-|z|/H),
\end{equation}
where
\begin{equation}
 H = (1+\alpha +\beta){a^2\over g_z}.
\end{equation}

\subsection{Linearized perturbations}
We consider ideal MHD equations written in lagrangian coordinates in the
rotating frame $({\bf x,y,z})$, centered at the distance $r_0$ from the
galactic center where the equilibrium velocity is ${\bf V}_o(r_0) =
r_0\Omega(r_0) {\bf y}$; $\Omega(r)$ and ${\bf V}_o(r)$ are the
equilibrium rotation frequency and azimuthal velocity. The frame is
defined by $y=r\theta,\ x=r-r_0$.\\
The displacement vector ${\bf \xi}$ and the perturbed velocity ${\bf v}$
satisfy the equation~:
\begin{equation}
({\p\over\p t}+ {\bf V}_o.{\bf\nabla}){\bf\xi}={\bf v}+ ({\bf\xi
.\nabla}) {\bf V}_o. \label{displacement}
\end{equation}
The induction equation, ${\bf\nabla}\times{\bf E}=-\p{\bf B}/\p t$ and
Ohm's law, ${\bf E}+{\bf v}\times {\bf B}={\bf 0}$ are easily found to
give the perturbed magnetic field:
\begin{equation}
{\bf b}={\bf\nabla}\times ({\bf\xi}\times {\bf B}_o), \nonumber
\end{equation}
and the continuity equation gives the perturbed density:
\begin{equation}
\rho=-\nabla.(\rho_o {\bf\xi}). \nonumber
\end{equation}
We look for perturbations of the form:
\begin{eqnarray}
{\bf\xi}(x,y,z,t)&=&\rho_o^{-1/2}(z)\;{\bf\tx}(x,y,t),\label{structz}\\
{\bf  v}(x,y,z,t)&=&\rho_o^{-1/2}(z)\;{\bf\tilde  v}(x,y,t).
\label{structvz}
\end{eqnarray}
As was shown by Parker (1967) an additional exp($ik_z z$) vertical
dependence can be added to the perturbations. This would  not change the
mathematical structure of the problem, and would still allow our
analytical treatment. Here we  restrict ourselves to the modes with a
vanishing vertical wavenumber, which are known to be the most unstable
ones.\\
The energy of these perturbations does not vanish at infinite height but
is bounded, thus the solution (\ref{structz})-(\ref{structvz}) allows a
simpler mathematical formulation with no loss of physical generality
(Lerche \& Parker 1967).\\

\section{Classical Parker instability}
To get the simplest formulation of the instability, we now consider a
magnetized gas layer with no cosmic rays and no rotation.\\
Writing Euler's equations after a Fourier transform in space
$(x,y)\rightarrow (k_x,k_y)$ and a Laplace transform in time
$t\rightarrow \omega$ leads to the usual dispersion equation, a
polynomial of degree 3 in $\omega^2$ (Parker 1967).  This dispersion
equation admits three different roots $\omega_{\rm S}^2,\omega_{\rm
A}^2,\omega_{\rm P}^2$ corresponding to the compresional magnetosonic
waves ($\omega_{\rm S}^2>0$), the torsional Alfven waves ($\omega_{\rm
A}^2>0$) and the ``Parker mode'' ($\omega_{\rm P}^2<\omega_{\rm A}^2$)
which is unstable if $\omega_{\rm P}^2<0$.\\
To each root $\omega_j^2(k_x^2)$ two polarizations correspond
($\xi_j(k_x)$ for $+\omega_j$ and $\xi_{j'}(k_x)$ for
$\omega_{j'}=-\omega_j$), and any initial perturbation projects onto the
basis of six polarizations $\xi_j(k_x)$.\\
The time evolution of an initial perturbation $\xi(k_x,t=0)$ is simply
the sum of the six independent, oscillating or exponential, time
evolutions of the projections  $P_j(k_x)$~:
\begin{equation}
\xi(k_x,t)=\sum P_j(k_x)\xi_j(k_x)
\e^{-i\omega_jt}.\label{decomp}
\end{equation}
The dispersion relation can be written in a dimensionless form~:
\begin{equation}
Q_3(\tom^2)-\bk_x^2 \Delta(\tom^2)=0, \label{dispersion}
\end{equation}
where $Q_3$ is a polynomial of degree $3$ of the convenient
dimensionless variable
\begin{equation}
\tom\equiv {H\omega\over V_{\rm A}}.\nonumber
\end{equation}
$V_{\rm A}=a (2\alpha)^{1/2}$ is the Alfven speed, and  $\bk$ is the
dimensionless wavenumber $Hk$.\\
The coefficients of $Q_3$ are independent of $k_x$, and are given in
Appendix~A.\\
$\Delta$ is a polynom of degree 2 in $\tom^2$, given by~:
\begin{equation}
\Delta(\tom^2)\equiv(1+2\alpha)\tom^4-2(1+\alpha)\left\{
\bk_y^2+{1\over4}\right\}\tom^2 +\bk_y^2 (\bk_y^2-\bk_{\rm P}^2),
\end{equation}
\begin{equation}
k_{\rm P}^2\equiv {1+\alpha\over2H^2}.
\end{equation}
The larger of its two roots, $\todp^2>\tod^2$, is always positive,
corresponding to stable Alfven waves. $k_{\rm P}$ appears as the maximum
azimuthal wavenumber allowed for Parker instability ($\tod^2<0$).
Perturbations with a wavelength shorter than $\lambda_{\rm
P}=2\pi/k_{\rm P}$ are stabilized by the high energetic  cost of bending
the magnetic field lines on short scales.\\
For $\bk_y<\bk_{\rm P}$, the negative root $\tod^2$ goes through a
minimum at $k_y=k_\Delta$ (see Fig.~\ref{fw_ky}). The growth of the
instability favours the azimuthal lengthscale
$\lambda_\Delta=2\pi/k_\Delta$ (increasing from $\sim 7 H $ to $\sim 12
H$ for $\alpha$ decreasing from 5 to 0.5), and the typical time scale of
the instability always ranges between once and twice the time scale
needed for an Alfven wave to cross the total height of the disk. We
define $T_{\rm Parker}$ as
\begin{equation}
T_{\rm Parker}\equiv {2H\over V_{\rm A}}. \nonumber
\end{equation}
In the limit of infinite $k_x$ the dispersion relation
(\ref{dispersion}) reduces to $\Delta(\tom^2)=0$ which we call the  {\it
asymptotic dispersion relation}.  We can approximate the unstable root
$\omega_{\rm P}(k_x^2)$ of the dispersion relation (\ref{dispersion})
when $k_x\to\infty$~:
\begin{equation}
{\omega_\Delta-\omega_{\rm P}\over\omega_\Delta}=
\left({\bk_\infty\over\bk_x}\right)^2+O({1\over\bk_x^4}),
\label{omdelta}
\end{equation}
where $\bk_\infty$ is defined, according to (\ref{dispersion}), as~:
\begin{equation}
\bk_\infty^2=-{Q_3(\tod^2)\over2\tod^2
{\p\Delta\over\p\tom^2}(\tod^2)}.\label{kinfini}
\end{equation}
A crude estimate leads to $\bk_\infty \le 1$ for $k_y=k_\Delta$.\\
The growth rate of the instability is maximum for infinite $k_x$ (Parker
1966). Stabilization by turbulence and diffusion at small scales would
in fact give an upper limit for $k_x$ (Zweibel \& Kulsrud 1975;
Lachi\`eze-Rey \etal 1980; Cesarsky 1980). The maximum $k_x$ is reached
when the growth time equals the diffusion time~: $\eta k_x^2\sim a/H$,
where $\eta$ is the diffusion coefficient of the magnetic field through
the gas. In HI regions, $\eta$ is determined by ambipolar diffusion, and
$k_{max}\sim 1pc^{-1}$; in ionized regions, the coefficient $\eta$ is
much smaller and the $k_{max}$ is several orders of magnitude larger.
These limits  are high enough, compared to the present scales ($\sim
1/H$), to consider  $\omega_\Delta$ as a good approximation of the most
unstable growth rate.\\
\begin{figure}
\picture 87.8mm by 63.8mm (fw_ky)
\caption[]{$\bk_y$-dependence of the growth rate at $\bk_x=\infty$ (full
line), and at $\bk_x=0$ without rotation (dashed line), and with
rotation (dotted line). $\alpha=\beta=1$. The cosmic rays are
responsible for the infinite slope at $\bk_y=0$ and $\bk_x=\infty$. A
higher CR pressure ($\beta>\alpha$) would destabilize the axisymmetric
mode}  \label{fw_ky}
\end{figure}
We derive in Appendix~A the eigenvector of the Parker mode, consistent
with the usual description of the instability mechanism~: as a loop
forms and moves up in the disk, it expands to adjust to the ambient
pressure, generating buoyancy forces which push the loop further upward.
The eigenvector shows the following features, typical of the instability
without rotation or cosmic rays~:
\par (i) We deduce from the relations of parity with respect to $k_x$
that a perturbation, initially symmetrical with respect to the ($y,z$)
plane ($\xi_x(k_x)$ odd while $\xi_y(k_x)$ and $\xi_z(k_x)$ are even),
will conserve its symmetry while expanding. The simplest example,
summarizing the relative $(x,y)$-symmetries of the components of the
Parker eigenvector, is the following~:
\begin{eqnarray}
\xi_x(x,y,t)=\Lambda_x \sin(k_xx)\cos(k_yy)\;\;\e^{-i\omega t}
\nonumber\\
\xi_y(x,y,t)=\Lambda_y\cos(k_xx)\sin(k_yy)\;\;\e^{-i\omega
t} \nonumber\\
\xi_z(x,y,t)=\Lambda_z\cos(k_xx)\cos(k_yy)\;\;\e^{-i\omega t} \nonumber
\end{eqnarray}
\par (ii) ($\Lambda_y/\Lambda_z$) is real, so the horizontal
displacement is maximum where the vertical displacement vanishes. Its
sign is the same as $k_y$, therefore the "crests" tend to widen along
${\bf y}$, and the "valleys" tend to narrow.
\par  (iii) ($\Lambda_x/\Lambda_z$) is real and the same sign as $k_x$.
The perturbed magnetic flux tubes, at a given $k_x$, are therefore
stretched along the ($x$)-direction at the crests, and are contracted
along the ($x$)-direction in the valleys.
\par  (iv) $(H\rho)/(\xi_z\rho_o)$ is real and negative, so density is
maximum in the valleys, and minimum at the crests.  This classical image
of gas draining down along the magnetic field lines is known to be a
misleading {\it explanation} for Parker instability, and should be
considered rather as a geometrical consequence of it~: while {\it both}
the thermal and magnetic pressures contribute to the vertical
equilibrium against the gravitation, the gas draining down the field
lines may gain some gravitational energy in exchange for some work of
the thermal pressure force {\it only}, taking advantage of the natural
anisotropy of the magnetic pressure (see Hughes \& Cattaneo 1987).\\
The most unstable perturbations have a polarization well fitted by~:
\begin{equation}
{\xi_x\over i\xi_z}\sim {k_1k_x\over k_x^2+k_2^2}\ll1.
\label{fit1}
\end{equation}
where $k_1<0$, and $k_2\sim k_y\sim1/H$.\\
This approximate formula outlines the scale $1/k_2\sim H$ of the natural
radial extension of the Parker instability.\\

\section{Effect of the cosmic-ray gas}
Although often neglected for simplicity, the pressure of the cosmic ray
gas is comparable to the thermal and magnetic pressure in the disk of
our galaxy ($P\sim 10^{-12}$ dyn.cm$^{-2}$).  Following Shu (1974), we
modelize cosmic rays as a gas whose pressure gradient is orthogonal to
magnetic field lines, leading to the usual picture~: gravity acts on the
gas, tied to  the magnetic field lines, while they respond  to the
cosmic ray pressure.\\
The linearized action of the CR pressure force on the gas is written in
terms of displacements~:
\begin{equation}
F_{\rm cr}\equiv {i\beta \rho_o a^2\over H} \left\{
\begin{array}{l}
-k_x\xi_z \\
-k_y\xi_z  \\
 k_x\xi_x+k_y\xi_y
\end{array}
\right.\nonumber
\end{equation}
This new force enters the Euler equations and still allows a projection
of any solution on a basis of six eigenvectors.

\subsection{Interchange-like instability}
We show in Appendix~A how the dispersion relation, especially the
asymptotic one ($\Delta(\tom^2)=0$), is modified~: setting $k_y=0$ gives
the only axisymmetric mode~:
\begin{equation}
\omega^2=(\alpha-\beta){1+\alpha+\beta\over 1+2\alpha}{a^2\over
H^2}\nonumber
\end{equation}
If the cosmic ray pressure exceeds the magnetic one ($\beta>\alpha$),
the CR gas destabilizes the axisymmetric mode into an interchange-like
instability though $\partial (B_o/\rho) /\partial z <0 $ (see Hughes \&
Cattaneo (1987), for a review of the differences between the undulate and
interchange magnetic instabilities).\\
The growth rate of this axisymmetric mode is larger than that of
Parker's undulate instability only if
\begin{equation}
{\beta\over\alpha}>3+4\alpha>3,\nonumber
\end{equation}
which is a very high ratio of CR to magnetic pressures. According to
observations (see the review by Boulares \& Cox 1990), the loosest
constraint would be $\alpha\sim 0.5 \mbox{ to } 5$ and $\beta\sim
0.5-5$.\\
The CR gas simply increases the strength of Parker's undulate
instability as long as $\beta /\alpha < 3+4\alpha$. For example, the
growth rate is multiplied by a factor $\sim 2$ from $\beta/\alpha=0$ to
$\beta/\alpha=1$ (see the expression of the optimum growth rate in
Appendix~A).

\subsection{Effect on the radial displacement}
The polarization of the unstable eigenvector is significantly modified
by the cosmic ray pressure (see Appendix~A).\\
The features (i) and (ii) of the preceding section are still valid.\\
However a rise of the regions of maximum density may now occur, at low
$k_x$, if the CR pressure is high enough, making the classical feature
(iv) only asymptotically valid ($|k_x|\gg k_{\infty}$).\\
The sign of the radial displacement depends on a new  critical azimuthal
wavenumber $k_{\rm cr}<k_{\rm P}$ defined as~:
\begin{equation}
k_{\rm cr}\equiv k_{\rm P}\left({\beta\over\alpha+\beta}{1+2\beta\over
2+2\beta}\right)^{1/2}\nonumber
\end{equation}
The feature (iii) is now replaced by~:
\par (iii') ($\Lambda_x/\Lambda_z$) is real and the same sign as
$(k_{\rm cr}-k_y)k_x$~: in the approximate formula (\ref{fit1}), the
sign of $k_1$ now depends on $k_y$.\\
If $k_y=k_{\rm cr}$, the radial displacement vanishes because of the
perfect radial equilibrium between the forces. The dispersion equation
is then satisfied by the asymptotic root $\tod$ whatever the radial
wavenumber $k_x$. \\
The most unstable asymptotic mode occurs at $k_y=k_\Delta$, and we
find~:
\begin{equation}
k_\Delta>k_{\rm cr} \Longleftrightarrow {\beta\over\alpha}<{2\over
1+(1+4\alpha)^{1/2}}\Leftrightarrow
\alpha>\beta(1+\beta)\nonumber
\end{equation}
This critical ratio of the CR to magnetic pressures always lies in the
$]0,1[$ interval (see Fig.~\ref{fwayrot}).\\
\begin{figure}
\picture 87.8mm by 63.8mm (fwayrot)
\caption[]{Radial extension of the magnetic flux tubes depending on the
cosmic ray pressure} \label{fwayrot}
\end{figure}
The cosmic ray
pressure rules the sign of the radial force as follows~: a low CR
pressure ($\alpha>\beta(1+\beta)$) favours an ($x,y$)-expansion of the
flux tubes on the "crests", and a compression in the "valleys". \\
A high CR pressure ($\alpha<\beta(1+\beta)$) leads to the inverse
picture as regard the $x$-direction (see Fig.~\ref{fwayrot}). The
expansion due to the CR pressure force is more efficient in the lower
regions where magnetic lines are vertically compressed than in the
higher regions where the magnetic lines are vertically diluted. Of
course this does not change the vertical expansion of the flux tube,
still allowing buoyancy force to act as usual.\\

\section{Effect of uniform rotation}
Shu (1974) and Zweibel and Kulsrud (1975) have considered the effect of
uniform rotation on the Parker instability; they essentially concluded
that the $k_y$-threshold of instability is unchanged.\\
We choose the gravitational potential so as to allow both the  uniform
rotation at equilibrium and the simple approximation that $g_z$ is
constant. The global mathematical formulation is not modified, so
solutions still project onto a basis of six eigenvectors. The
polarization of the eigenvectors and the polynomial $Q_3(\tom^2)$ are
modified. Rotation introduces a new dimensionless parameter
$\tilde\Omega=\Omega T_{Parker}$ which can be estimated with the
convenient formula~:
\begin{equation}
\tilde\Omega\equiv{2\Omega H\over V_{\rm A}}\sim 12.3{H({\rm pc})\over
T_\Omega(10^6{\rm yr})V_{\rm A}({\rm
km.s}^{-1})}\nonumber
\end{equation}
If the vertical gravitation field were mainly generated by the
gravitation of the central part of the galaxy, the scale height of the
disk would depend on the rotation frequency as~:
\begin{equation}
H^2=(1+\alpha+\beta){2a^2\over\Omega^2}\nonumber
\end{equation}
and would lead to a scale-height $\sim 5$ times higher than the one
observed, because this estimation omits the contribution of the
self-gravity of the disk of stars $+$ gas in confining the galactic
gas.\\
The rotation parameter is therefore always $>2$ when self gravity is
neglected, and is $\sim 0.7$ according to the observed values in the
vicinity of the sun. It is not negligible in either case.\\
In the rotating frame, two inertial forces now act on the perturbed
gas~:
\par  (i) the centrifugal force $\rho_o\Omega^2\xi_x{\bf x}$ cancels out
exactly the radial gravitational force if rotation at equilibrium is
uniform.
\par (ii) the Coriolis force is equal to $2\rho_o{\bf v\times \Omega}$
and couples the azimuthal displacement to the radial ones. The gas tends
to move away from the galactic centre where $v_y>0$, and tends to fall
towards the galactic centre where $v_y<0$, while sliding down along the
magnetic field lines.\\
Hence rotation breaks the symmetry with respect to the ($y,z$) plane.
Although the roots of the dispersion relation $\omega(k_x^2)$ are still
even with respect to $k_x$, the radial displacement $\xi_x$ no longer
vanishes at $k_x=0$ (see the polarizations of the eigenvectors in
Appendix~A)~:
\begin{equation}
{\xi_x\over\xi_y}(\bk_x=0)={-i\tom\tilde\Omega\over
\bk_y^2-\tom^2},\nonumber
\end{equation}
which is strictly positive for the unstable Parker mode  ($-i\tom_{\rm
P}~>~0$). Rotation consequently excludes the possibility of a radial
equilibrium independent of $k_x$. The previous contributions to $\xi_x$
were odd in $k_x$, and thus odd in $x$ for a perturbation that is even:
this corresponds to the radial expansion or compression of the flux
tubes; the new contribution due to rotation is even in $k_x$: it
corresponds to a twisting of the flux tubes, which now become
helical.\\
The radial stretching of the field lines due to the Coriolis force
enhances the vertical magnetic tension and consequently diminishes the
growth rate of the instability (see Fig.~\ref{fw_ky}).\\
The radial equilibrium between the Coriolis force and the combination of
pressure and tension forces occurs at a unique $k_x=k_o$ given by~:
\begin{equation}
Hk_o=\tilde\Omega f_{o}(\alpha,\beta,\bk_y),\nonumber
\end{equation}
where $f_{o}$ is independent of $\Omega$ and accounts for the mixed
contributions of magnetic and thermal radial pressures, and of the
magnetic tension. $k_o$ is negative if $k_y<k_{\rm cr}$ and positive if
$k_{\rm Q}>k_y>k_{\rm cr}$ (see Appendix~A). \\
The response of the Parker instability to a radially localized
perturbation will consequently loose its initial symmetry, giving  a
clockwise helix around the azimuthal direction. Nevertheless, the radial
extension of the helix remains very small compared to its vertical size
as far as the most unstable wavenumbers are concerned.\\
Indeed, the effect of rotation on the direction of the radial
displacement becomes negligible when $|k_x/k_o|\gg 1$, because radial
pressure forces then overwhelm the Coriolis force. The asymptotic
dispersion equation $\Delta(\tom^2)=0$ as well as the estimate
(\ref{omdelta}) are unaffected by rotation (a crude estimate now leads
to  $\bk_{\infty}\le 1+\tilde\Omega$), although $k_o$ can be arbitrarily
large (it is infinite if $k_y=k_{\rm cr}$).\\
Before introducing differential rotation, let us recall the six
polarizations of the three physically distinct modes of evolution, in
the asymptotic limit $|\bk_x|\to\infty$.\\
The asymptotic polarization of the Parker mode $\xi_{\rm P}(k_x)$ is,
when $|\bk_x|\to \infty$, essentially in the ($x,y$)-plane~:
\begin{eqnarray}
{\xi_{{\rm P}z}(k_x)\over\xi_{{\rm P}y}(k_x)}&\sim&-{i\over \bk_y}
{(1+2\alpha)\tod^2-\bk_y^2 \over
1+\alpha+\beta}+O\left({1\over\bk_x}\right),\label{polyzy}\\
{\xi_{{\rm P}x}(k_x)\over \xi_{{\rm P}y}(k_x)}&\equiv&{-1\over
2\bk_x\bk_y} {(1+2\beta)\tod^2+\bk_y^2 \over
1+\alpha+\beta}+O\left({1\over\bk_x^2}\right).\label{polyxy}
\end{eqnarray}
The asymptotic polarization of stable Alfven waves $\xi_{\rm A}(k_x)$ is
given by the same equation, replacing $\tod$ with  $\todp$. The
polarizations associated with the opposite frequencies are identical~:
$\xi_{\rm P'}(k_x)=\xi_{\rm P}(k_x)$ and $\xi_{\rm A'}(k_x)=\xi_{\rm
A}(k_x)$.\\
The asymptotic polarization of the magnetosonic waves, associated with
the diverging frequency $\tom_{\rm S}\sim (1+(2\alpha)^{-1})^{1/2} k_x$
when $|\bk_x|\to \infty$, depends on the strength of rotation, but is
essentially radial~:
\begin{eqnarray}
{\xi_{{\rm S}y}(k_x)\over\xi_{{\rm S}x}(k_x)}&\sim&{\bk_y-
i\tilde\Omega(2\alpha(1+2\alpha))^{1/2} \over
(1+2\alpha)\bk_x}+O\left({1\over\bk_x^2}\right),\nonumber\\
{\xi_{{\rm S}z}(k_x)\over \xi_{{\rm S}x}(k_x)}&\sim &{i\over
2\bk_x}{1+2\beta\over 1+2\alpha}  +O\left({1\over\bk_x^2}\right).
\nonumber
\end{eqnarray}
The polarization $\xi_{\rm S'}(k_x)$ of a magnetosonic wave with an
opposite frequency $\tom_{\rm S'}=-\tom_{\rm S}$ is obtained by changing
the sign of $\tilde\Omega$ in the polarization of $\xi_{\rm S}(k_x)$.\\

\section{Modelization of differential rotation}

\subsection{Shearing-sheet equations}
We use the standard "shearing-sheet" modelization of differential
rotation (Goldreich \& Lynden-Bell 1965), extensively used in the theory
of spiral density waves. In this model one sticks to the minimal physics
of differential rotation, ignoring all radial variations of equilibrium
quantities except the rotation frequency, and neglecting the effects of
cylindrical geometry.\\
Without rotation the instability has a purely imaginary frequency, \ie
it is stationary. With solid rotation the frequency has a real part
allowing the mode to be stationary in the moving frame. With
differential rotation it can be stationary only at one radius (the
corotation radius).\\
We work in the frame rotating at the corresponding frequency, so that
the total derivative with respect to time is~:
\begin{equation}
{\d\over \d t}\equiv{\partial\over\partial t}+2iAk_yx, \nonumber
\end{equation}
where $x$ is still the local radial coordinate and $A$ is the Oort
constant defined at the corotation radius $r_{0}$ by~:
\begin{equation}
A\equiv {r_{0}\over 2}{\partial\Omega\over\partial r}<0. \nonumber
\end{equation}
The definition (\ref{displacement}) of the lagrangian displacement as in
Sect.~2, leads to
\begin{eqnarray}
v_x&=&{\d\xi_x\over \d t}\nonumber\\
v_y&=&{\d\xi_y\over \d t}-2A\xi_x\nonumber\\
v_z&=&{\d\xi_z\over \d t} \nonumber
\end{eqnarray}
The shearing-sheet approximation leads to the following system of
equations~:
\begin{eqnarray}
{\d^2\xi_x\over \d t^2}&=&+2\Omega{\d\xi_y\over \d t}-4A\Omega\xi_x+
{F_x\over\rho}\label{Eulerx}\\
{\d^2\xi_y\over \d t^2}&=&-2\Omega{\d\xi_x\over \d t}+{F_y\over\rho}
\label{Eulery}\\
{\d^2\xi_z\over \d t^2}&=& +{F_z\over\rho}\label{Eulerz}
\end{eqnarray}
where ${\bf F}$ stands for the total sum of the thermal, magnetic and CR
pressure forces, the gravitation and the magnetic tension forces.\\
This formulation in terms of lagrangian displacements outlines the two
usual inertial forces existing in any rotating frame with angular speed
$\Omega_{(r_{0})}$~:
\par  (i) The Coriolis force has already been
introduced in Sect.~5.
\par  (ii) The new radial force here is the sum of the centrifugal
force
\begin{equation}
F_{\rm cent.}(x)=\rho_o \Omega^2_{(r_{0})}(r_{0}+x), \nonumber
\end{equation}
and the radial gravitational force given by the radial hydrostatic
equilibrium
\begin{equation}
g_r(x)=-\Omega^2_{(r_{0}+x)}(r_{0}+x). \nonumber
\end{equation}
This sum used to exactly vanish in a disk with uniform rotation.\\
The resulting radial differential force
\begin{equation}
F_{\rm cent.}(\xi_x)+\rho_o g_r(\xi_x)\sim-4A\Omega\rho_o\xi_x\equiv
\rho_o r_{0}{\p\over\p r}\left({g_r\over r}\right)\xi_x \nonumber
\end{equation}
is destabilizing ($ A\Omega<0$), but the Coriolis force stabilizes the
perturbations into epicycles of frequency $\kappa$, with~:
\begin{equation}
\kappa^2 \equiv 4\Omega(\Omega+A). \nonumber
\end{equation}
This stabilization occurs as long as shear is reasonable
(i.e.~$-A/\Omega<1$), which is true along the observed rotation curve of
spiral galaxies~: rotation is uniform near the galactic center
($-A/\Omega\sim 0$), becomes Keplerian away from the bulge
($-A/\Omega\sim 3/4$), and finally flattens in the external parts
($-A/\Omega\sim 1/2$).\\
The standard separation of the vertical and horizontal structures,
(\ref{structz})-(\ref{structvz}), is still possible. Performing a radial
Fourier transform then leads to define $\hat\xi(k_x,t)$ as~:
\begin{eqnarray}
\xi(x,z,t)=\rho_o^{-1/2}(z)\int_{-\infty}^{+\infty} \e^{ik_xx}
\hat\xi(k_x,t)\d k_x,\label{Four1}
\end{eqnarray}
The total derivative in Fourier space becomes
\begin{eqnarray}
{d\over dt}&\equiv & {\p\over\p t}-2Ak_y{\p\over\p k_x}.
\end{eqnarray}
The differential system satisfied by $\hat\xi$ can be written in the
following vector form~:
\begin{equation}
{\d^2{\bf\hat\xi}\over \d t^2}=-2{\bf\Omega}\times{\d{\bf\hat\xi}\over
\d t}-{a^2\over H^2} {\cal L}(\bk_x)[{\bf\hat\xi}],\label{systdiff1}
\end{equation}
where the dimensionless hermitian matrix ${\cal L}(\bk_x)$, accounting
for the contributions of the various forces, {\it  including the
differential force}, is defined as~:
\begin{equation}
\left(
\begin{array}{ccc}
\displaystyle{\bk_x^2+2\alpha(\bk_x^2+\bk_y^2-\bk_A^2)}&
\displaystyle{\;\;\;\;\;\bk_x\bk_y\;\;\;\;\;}& \displaystyle{i({1\over
2}+\beta)\bk_x}\\
 & & \\ \times & \displaystyle{\bk_y^2 }&
\displaystyle{i({1\over2}+\alpha+\beta)\bk_y}\\
 & & \\ \times & \times & \displaystyle{{1+2\alpha\over
4}+2\alpha\bk_y^2}
\end{array}
\right) \nonumber
\end{equation}
We have introduced the dimensionless strength of the differential
force
\begin{equation}
\bk_A^2\equiv -{4A\Omega H^2\over V_{\rm A}^2}>0. \nonumber
\end{equation}

\subsection{Two equivalent formulations}
Two equivalent approaches may be adopted to solve this differential
system. Both lead to a sixth order system of ordinary differential
equations, depending on the radial wavenumber in the  first case, and on
time in the second. Both are helpful in understanding how differential
rotation couples the radial Fourier-variable ($k_x$) to the
temporal-variable ($t$).\\

\subsubsection{The $k_x$-formulation}
Following the method of Lin \& Thurstans (1984), we can separate
variables by projecting $\hat\xi$ onto a basis of six independent
functions $\check\xi_j$~:
\begin{equation}
\hat\xi(k_x,t)=\sum P_j(k_x+2Ak_yt)\check\xi_j(k_x),\nonumber
\end{equation}
where the six scalar functions $P_j$ are arbitrarily chosen according to
the kind of perturbations studied, and the vectorial functions
$\check\xi_j$  are six independent solutions of
\begin{equation}
4A^2\bk_y^2{\p^2{\bf\check\xi}\over\p\bk_x^2}=4A\bk_y{\bf\Omega}\times
{\p{\bf\check\xi}\over \p \bk_x}-{a^2\over H^2}{\cal
L}(\bk_x)[{\bf\check\xi}]. \label{systdiff2}
\end{equation}
The general evolution of any perturbation is consequently written as~:
\begin{eqnarray}
\xi(x,z,t)=\sum\rho_o^{-1/2}(z)\int_{-\infty}^{+\infty}
\e^{ik_xx} P_j(k_x+2Ak_yt)\check\xi_j(k_x)\d k_x.\nonumber
\end{eqnarray}
For each polarization $\check\xi_j$, the choice of the arbitrary
function $P_j$ allows us to describe standing  waves amplified
exponentially with time as well as transient wave packets.\\
In both cases the scalar functions $P_j$ depend only on the Fourier
transform $\hat\xi(k_x,t=0)$ of the initial radial shape of the
perturbation~:
\begin{equation}
\hat\xi(k_x,t=0)=\sum P_j(k_x)\check\xi_j(k_x),\label{initialshape}
\end{equation}
In particular, the response to a radially localized perturbation (a
constant in the $k_x$-Fourier space) is studied in Sect.~6.6.\\
This formulation is not well adapted to  describe the transition from
$A=0$ to $A\neq 0$. Thus the analytical  work will also use the
equivalent temporal formulation described in the next subsection.

\subsubsection{The temporal formulation}
Let us define
\begin{equation}
\tx(k_x,t)=\hat\xi(k_x-2Ak_yt,t). \label{tildexi}
\end{equation}
$\tx$ must satisfy the following differential system~:
\begin{equation}
{\p^2{\bf\tx}\over \p t^2}=-2{\bf\Omega}\times{\p{\bf\tx}\over \p
t}-{a^2\over H^2} {\cal L}(\bk_x-2A\bk_yt)[{\bf\tx}].\label{systdiff3}
\end{equation}
The
general evolution of any perturbation is consequently written as~:
\begin{equation}
\xi(x,z,t)=\rho_o^{-1/2}(z)\int_{-\infty}^{+\infty}
\e^{i(k_x-2Ak_yt)x} \tx(k_x,t)\d k_x.\label{Fourierinv}
\end{equation}
This formulation allows us to outline the limit of uniform rotation~: if
$A=0$, $\tx$ identifies with the real Fourier transform $\hat\xi$ which
therefore satisfies the classical differential system with
time-independent coefficients~:
\begin{equation}
{\p^2{\bf\hat\xi}\over\p t^2}=-2{\bf\Omega}\times {\p{\bf\hat\xi}
\over\p t}-{a^2\over H^2}{\cal L}(\bk_x) [{\bf\hat\xi}], \nonumber
\end{equation}
which was studied in Sect.~5.\\
Following this approach, we can characterize the presence of
differential rotation with the new time-scale $T_{\rm Shear}$ defined
as
\begin{equation}
T_{\rm Shear}\equiv -{1\over A}, \nonumber
\end{equation}
intervening both~:
\par (i) as the typical time-scale of the linear growth of the radial
wavenumber
\begin{equation}
K_x(t)\equiv k_x-2Ak_yt, \label{klineaire}
\end{equation}
\par (ii) as a scaling of the new differential force compared to the
only other radial force remaining at $\bar K_x(t)=0$
\begin{equation}
-{4A\Omega\over k_y^2V_{\rm A}^2}={\bk_A^2\over k_y^2}
\sim{2\pi\over\bk_y^2}{T_{\rm Parker}^2\over T_\Omega T_{\rm
Shear}}.\label{scaling}
\end{equation}
In order to keep apparent the limit of vanishing shear, we have choosen
to study analytically the function $\tx(k_x,t)$ in the following
sections.\\
The correspondence between the two formulations is simply given by
\begin{equation}
\tx(k_x,t)=\sum P_j(k_x)\check\xi_j(k_x-2Ak_yt). \label{corresp}
\end{equation}

\section{Numerical simulations}
In order to compute the effect of differential rotation on the Parker
instability, we perform numerical simulations,  using an implicit
Runge-Kutta method to integrate the differential system
(\ref{systdiff3}).\\
We solve here the time-evolution $\xi_{k_{0x}}(t)$ of a perturbation
whose initial radial structure is a wave of definite wavenumbers
$(k_{0x},k_y)$, and whose polarization is essentially along the Parker
unstable asymptotic mode ($\xi_{\rm P}(k_{0x})$ with $k_y<k_{\rm P}$ and
$k_{0x}\ll -k_{\infty}$). Apart from some marginal cases, the evolution
of any other initial polarization is anyhow dominated by the unstable
Parker mode after a few growth times, as will be shown analytically in
Sect.~8.
\begin{eqnarray}
P_j(k_x)&=&\delta(k_x-k_{0x})\nonumber\\
K_x(t)&=&k_{0x}-2Ak_yt\nonumber\\
\xi(x,z,t)&=&\rho_o^{-1/2}(z)\e^{iK_x(t)x}\check\xi(K_x(t))\nonumber
\end{eqnarray}
Confirming the result of Shu (1974), the Parker instability occurs in
all the simulations, at least asymptotically ($t\to+\infty$), for all
$\bk_y<\bk_{\rm P}$. The asymptotic growth  rate appears to be $|\od|$,
for $|\bar K_x(t)|>\bk_\infty$. At high $|K_x|$, the effect of the
inertial forces (the differential force as well as the Coriolis force)
is negligible compared to the other forces so that this asymptotic
growth rate does not depend on the rotation parameters ($A,\Omega$). An
original demonstration of this asymptotic behaviour is given in
Sect.~8.2.\\
The behaviour of the solution at low $|\bar K_x(t)|$ is much more
complex, depending on three time-scales $T_{\rm Parker},T_\Omega,T_{\rm
Shear}$.\\
We illustrate, with numerical simulations, four different behaviours,
which are approached analytically, in the limit of weak  differential
rotation, in Sect.~8.\\
We can notice a strong similarity with the simulations performed by Ryu
\& Goodman (1992) in order to study the convective instability in
accretion disks. Although the nature of their instability is  different,
various effects described below are also found in their  numerical
results.\\

\subsection{Classical adiabatic growth}
\begin{figure}
\picture 87.8mm by 63.8mm (fclassicadiabatic)
\caption[]{The classical Parker instability at $k_y=k_\Delta<k_{\rm Q}$
when differential rotation is extremely low ($-A/\Omega=0.02$). The
logarithm of the radial (x), azimuthal (y) and vertical (z) displacement
$\xi$ are represented here. The time dependent radial wavenumber
$K_x(t)$ is in $H^{-1}$ units. The Coriolis force is responsible for the
slower growth at low $|K_x(t)|$, and for the change of sign of $\xi_x$
at $\bar K_x= \tilde\Omega f_o(1,0,\bk_\Delta)\sim 5.17$. Due  to the
extremely low shear, the integration is performed over a very long
time, explaining the very large final amplification obtained}
\label{fclassicadiabatic}
\end{figure}
If the shear is low enough, the differential force will be negligible
compared to all the other forces. The effect of shear is then reduced to
a slow linear growth of the radial wavenumber with time given by
Eq.~(\ref{klineaire}).\\
The slow growth of $K_x$ allows an adiabatic evolution of the solution~:
the instantaneous growth rate and polarization of the solution identify
with the growth rate and polarization found for uniform rotation
(Sect.~5), where $k_x$ is replaced with $K_x(t)$.\\
Hence it is not surprising that when $k_y<k_{\rm Q}$, the growth rate
diminishes significantly at low $K_x(t)$ (Coriolis force effect), and
that $\tx_x(t)$ changes its sign  at $K_x(t)=k_o$ (see Fig.
\ref{fclassicadiabatic}).\\
Besides, we can check that when $k_{\rm Q}<k_y<k_{\rm P}$, the growth
rate decreases, vanishes, and changes the Parker instable mode into an
oscillatory mode at low $K_x$. The lower the shear, the longer the time
spent at low $K_x$, and the higher the number of oscillations. This
range of azimuthal wavenumbers nevertheless does not lead to the most
unstable asymptotic growth rate $|\tod|$ ($k_\Delta<k_{\rm Q}$), so that
such oscillations are not strongly favoured (see Fig~\ref{fw_ky}).\\

\subsection{Transient adiabatic stabilization}
\begin{figure}
\picture 87.8mm by 63.8mm (ftransientadiabatic)
\caption[]{Transient adiabatic stabilization by differential rotation.
$-A/\Omega=0.2, \alpha=1, \beta=0, k_y=k_\Delta$}
\label{ftransientadiabatic}
\end{figure}
Both the Coriolis force and the differential force have a stabilizing
influence on the Parker instability at low $|K_x(t)|$, by enhancing the
radial excursions of the field lines, but the nature of their action is
different~:
\par (i) The Coriolis force (see the effect uniform rotation in Sect.~5)
couples the radial displacements to the azimuthal ones~: though it
significantly diminishes the growth rate of the instability, it is
unable to really stabilize it.
\par (ii) A strong differential force changes the radially stabilizing
magnetic tension into a radially destabilizing force ($\bk_A>\bk_y$),
which stretches the whole field line away from corotation. This radial
excursion of the field line changes the Parker instability into a
transient oscillatory mode (see  Fig.~\ref{ftransientadiabatic}).\\
In order to isolate this effect, the parameters have been chosen to allow
both an adiabatic evolution
$$T_{\rm Parker}\ll T_{\rm Shear}\Longleftrightarrow
\left|{A\over\Omega}\right| \ll {a\over \Omega H}\alpha^{1/2},$$
and let the differential force dominate the radial magnetic tension
$$\bk_A^2\ge \bk_y^2\Longleftrightarrow \alpha\le
{2\over\bk_y^2}\left|{A\over \Omega}\right|\left({\Omega H\over
a}\right)^2.$$
These two conditions can be fulfilled when both the differential
rotation and the magnetic field are weak.

\subsection{Non-adiabatic stabilization and wave emission}
\begin{figure}
\picture 87.8mm by 63.8mm (fwave)
\caption[]{Emission of Alfven and magnetosonic waves at realistic shear
($-A/\Omega=0.4, \alpha=1, \beta=0$) during the evolution of the Parker
instability. The magnetosonic waves appear at  $K_x>0$, mainly on
$\xi_x$ because their polarization at high $|K_x|$ is essentially
radial (see Sect.~5)}  \label{fwave}
\end{figure}
When the shear time is comparable with the Parker time, we find a strong
linear coupling between the unstable Parker mode and the oscillatory
magnetosonic and Alfvenic modes: these waves are emitted as  the Parker
mode passes at low $K_x(t)$. The magnetosonic oscillations are easily
identified~: their frequency increases with $K_x$, and the displacements
are essentially radial.\\
The Alfvenic oscillations are much slower, and are visible on the
vertical and azimuthal displacements.\\
These oscillations go together with a partial damping of the Parker mode
(see Fig.~\ref{fwave}), which is a combination of a shearing
stabilization (outlined in Sect.~8.3.3) and the energetic cost of wave
emission.\\

\subsection{Turn-over}
\begin{figure}
\picture 87.8mm by 63.8mm (fturnover)
\caption[]{If shear is high enough (here the rotation is ``flat''~;
$-A/\Omega=0.5$ with $\alpha=1, \beta=0, k_y=k_\Delta$), a turn-over
occurs at low $|K_x|$: together with the linear coupling to the Alfvenic
and magnetosonic waves, the vertical displacement changes sign. Here we
use a linear scale to better  show the change in the sign of $\xi_z$ and
$\xi_y$}  \label{fturnover}
\end{figure}
When the shear rate ($-A/\Omega<1$) exceeds a critical value, the
oscillations may lead to a turn-over~: the vertical motion of the gas,
which is the essence of the Parker instability, may be found to reverse
its direction, \ie gas which had started at $K_x<-\bk_\infty$ moving
upward may  be found at $\bar K_x>\bk_\infty$ to sink towards the
midplane (see Fig.~\ref{fturnover}).\\
This feature may lead to surprising inversion effects in the case of
radially localized perturbations (see Sect.~8.5).\\

\section{Analytical study}

\subsection{Mathematical transformation of the differential system}
We start from the function ${\bf\tx}(k_x,t)$, defined in
Eq.~(\ref{tildexi}), whose asymptotic temporal growth is assumed to be
at most exponential, and we perform  a Laplace transform:
\begin{equation}
\bx(k_x,\omega)=\e^{-{i\omega k_x\over2Ak_y}} \int_0^{+\infty}
\e^{i\omega t}\tx(k_x,t)\d t.\label{Laplace}
\end{equation}
After some algebraic calculations, detailed in Appendix~B, using the
differential system (\ref{systdiff3}), we find that each of the three
functions ($\bx_x,\bx_y,\bx_z$) satisfies a second order ordinary
differential equation of the form:
\begin{equation}
\left\{{\p^2\over\p\tom^2} +{\p\log f(\tom)\over\p\tom}{\p\over\p\tom}
+g(\tom) \right\}\bx(k_x,\tom) =h(k_x,\tom),\label{Frobenius}
\end{equation}
where $h$ depends linearly on the initial conditions, and ($f,g,h$) are
singular where $\Delta(\tom^2)=0$ (see Appendix~B).\\
The space of solutions of Eq.~(\ref{Frobenius}) is two-dimensional; the
sixth-order character of the  problem is present here only through the
initial conditions, on the  right-hand side. Furthermore, $\hat\xi$ is
the only solution which is  bounded when $Re(\tom)\to\pm\infty$, so as
to allow an inverse Laplace transform~:
\begin{equation}
\tx(k_x,t)={1\over 2\pi}\int_{ip-\infty}^{ip+\infty}\e^{-i\omega
\left(t-{k_x\over2Ak_y}\right)}\bx(k_x,\omega)\d\omega,
\label{invLaplace}
\end{equation}
where $p$ is real and larger than the largest imaginary part of the
singularities of $\bx$.\\

\subsection{Asymptotic time-behaviour of the solution}

\subsubsection{Identification of the asymptotic time-behaviour on the
properties of the Laplace transform}
When $(t-k_x/2Ak_y)\to +\infty$, the major contributions to the inverse
Laplace integral (\ref{invLaplace}) can be {\it localized} in the
complex plane by properly moving the integration contour (see
Fig.~\ref{fcontour}), and can be decomposed into two distinct
contributions, of two saddle points and four singularities~:
\begin{equation}
\tx(k_x,t)=\tx_{\rm sad.}(k_x,t)+\tx_{\rm sing.}(k_x,t).
\end{equation}
\par (i)  As in the theory of spiral density waves (Lin and Thurstans
1984), the behaviour of $\bx$ when Re($\tom)\to\pm\infty$ reveals two
saddle point contributions (see Sect.~8.3). These contributions
correspond to the compressive magnetosonic waves propagating radially.
The WKB approximation, developed in Sect.~8.3, allows us to write, when
$K_x(t)\to+\infty$~:
\begin{equation}
\tx_{\rm sad.} (k_x,t)\sim\mu_{+}(k_x,\tom_{\rm
S})\e^{i\Psi_{+}(\tom_{\rm S})}+\mu_{-}(k_x,-\tom_{\rm
S})\e^{i\Psi_{-}(-\tom_{\rm S})},
\end{equation}
where $\mu_{+}(k_x,\tom_{\rm S}),\mu_{-}(k_x,-\tom_{\rm
S})$ are the magnetosonic  polarizations associated to the magnetosonic
frequencies $\pm\omega_{\rm S}(t)$, diverging asymptotically as
$K_x(t)\to+\infty$ like~:
\begin{equation}
\omega_{\rm S}(t)\sim (a^2+V_{\rm A}^2)^{1/2}K_x(t).
\end{equation}
$\Psi_{\pm}(\pm\tom_{\rm S}(t))$ is the characteristic magnetosonic
phase, described in more detail in Sect.~8.3.
\par (ii) We show in Appendix~C that the only true singularities of
$\bx(k_x,\tom)$ are the 4 complex roots $\tom_j\in\{\pm\tod,\pm\todp\}$
of the asymptotic dispersion relation $\Delta(\tom^2)=0$. These
singularities must be bypassed when moving the integration contour of
(\ref{invLaplace}), giving birth to 4 purely magnetic contributions.\\
A local study of the solutions of (\ref{Frobenius}), using the Frobenius
method, allows us to give the following expansions in the vicinity of
$\tom_j$~:
\begin{eqnarray}
\bx_x(k_x,\tom)&=&\Lambda_x^j(k_x)\bx_{0x}^j(\tom)\ln(\tom-\tom_j)
+\bx_{{\rm reg}_x}^j\nonumber\\
\bx_y(k_x,\tom)&=&{\Lambda_y^j(k_x)\over\tom-\tom_j}+
\Gamma_y^j(k_x)\bx_{0y}^j(\tom)\ln(\tom-\tom_j)+\bx_{{\rm reg}_y}^j
\nonumber\\ \bx_z(k_x,\tom)&=&{\Lambda_z^j(k_x)\over\tom-\tom_j}
+\Gamma_z^j(k_x)\bx_{0z}^j(\tom)\ln(\tom-\tom_j)+\bx_{{\rm reg}_z}^j
\nonumber
\end{eqnarray}
where $\bx_{\rm reg}^j(k_x,\tom)$ is regular at $\tom_j$, and $\bx_0^j$
is the  unique solution of the homogeneous equation associated with
Eq.~(\ref{Frobenius}) which is regular at $\tom_j$, normalized as
$\bx_0^j(\tom_j)=1$ (see Appendix~C).\\
Each singularity adds a contribution $\tx{\rm sing_j}(k_x,t)$ to the
inverse Laplace integral when $t\to +\infty$, which we can write as~:
\begin{equation}
\tx{\rm sing_j}(k_x,t)\sim \e^{-i\omega_j (t-{k_x\over2Ak_y})} \left|
\begin{array}{c} \Lambda_x^j(k_x)(H/V_{\rm A}t)+O(1/t^2)\\
\Lambda_y^j(k_x)+O(1/t)\\ \Lambda_z^j(k_x)+O(1/t)
\end{array}
\right.
\end{equation}
By inserting this asymptotic behaviour in the differential system
(\ref{systdiff3}), and considering the leading terms only, we find that
the asymptotic polarization $[\Lambda_x^j(k_x)(H/V_{\rm
A}t),\Lambda_y^j(k_x),\Lambda_z^j(k_x)]$ is necessarily parallel to the
corresponding asymptotic polarization $\xi_j(k_x)$ described in Sect.~5
($j=$ A, A', P or P').  Choosing the normalization $\xi_{j_y}(k_x)=1$,
we can write the global contribution of the four singularities as~:
\begin{eqnarray}
\tx_{\rm sing.} (k_x,t)&\sim&  P_{\rm A}(k_x)\xi_{\rm
A}(K_x(t))\e^{+i\omega_{\delta}{K_x(t)\over2Ak_y}}\nonumber\\ &+&P_{\rm
A'}(k_x)\xi_{\rm
A'}(K_x(t))\e^{-i\omega_{\delta}{K_x(t)\over2Ak_y}}\nonumber\\ &+&P_{\rm
P}(k_x)\xi_{\rm
P}(K_x(t))\e^{+i\omega_{\Delta}{K_x(t)\over2Ak_y}}\nonumber\\ &+&P_{\rm
P'}(k_x)\xi_{\rm P'}(K_x(t))\e^{-i\omega_{\Delta}{K_x(t)\over2Ak_y}}
\nonumber
\end{eqnarray}
We must distinguish here between two cases, according to the stability
of the system.
\par (i) if $k_y>k_{\rm P}$, the 4 roots $\tom_j$ are real, and
correspond to slow ($\tod$) and fast ($\todp$) Alfvenic oscillations.
All the solutions of the differential system (\ref{systdiff3}) are
bounded in this situation.
\par (ii) if $k_y<k_{\rm P}$, 2 roots are real ($\todp^2>0$),
corresponding to the  Alfven waves, and 2 roots are complex conjugate
($\tod^2<0$), corresponding to the unstable  (Im$[\tod]>0$) and damped
(Im$[\tod]<0$) Parker  solutions. The solution of the differential system
(\ref{systdiff3}) is consequently dominated, when $t\to +\infty$, by
this divergent contribution to the inverse Laplace integral
(\ref{invLaplace}).\\
This confirms the result of Shu (1974) concerning the impossibility of a
global stabilization by differential rotation.\\
\begin{figure}
\scaledpicture 203mm by 161mm (fcontour scaled 430)
\caption[]{Integration contour in the complex plane for the inverse
Laplace transform. The contour is moved in order to cross the saddle
points, and bypass the singularities. In dotted lines, two cuts in the
comlex plane link the four singularities~:$+\tod$ to $-\tod$, and
$+\todp$ to $-\todp$. In dashed lines, the positions of the saddle
points change with $K_x(t)$, ranging from $\tom_{\rm P_0}$ to $\tod$
(Parker saddle), from $\tom_{\rm A_0}$ to $\todp$ (Alfven saddle), and
from $\tom_{\rm S_0}$ to real infinity (magnetosonic saddle)}
\label{fcontour}
\end{figure}

\subsubsection{Definition of the complex constants ${\cal C}_{mn}$
coupling the asymptotic states}
Following (\ref{corresp}), the Laplace transform can equivalently be
related to $\cx(k_x)$ as
\begin{equation}
\bx(k_x,\omega)={-1\over 2Ak_y}\sum P_j(k_x)\int_{k_x}^{+\infty}
\e^{-{i\omega K_x\over 2Ak_y} }\cx(K_x)\d K_x.\label{Laplace2}
\end{equation}
The {\it localization} of the contributions to the inverse Laplace
transform permits us to separate and identify six asymptotic modes of
evolution when $K_x(t)\to+\infty$. For convenience, we use the equivalent
$k_x$-formulation of Sect.~6.2.1 and define for each mode (noted
$m\in\{{\rm A, A', P, P', S, S'}\}$) a function $\cx^m(k_x)$, solutions
of (\ref{systdiff2}) whose asymptotic behaviour when $k_x\to +\infty$
identifies with this mode.\\
Besides, we can read on the differential system (\ref{systdiff2}) that
if $(\cx_x,\cx_y,\cx_z)(k_x)$ is a solution, then
$(\cx_x,-\cx_y,-\cx_z)(-k_x)$ is a solution too. Consequently, the
separation and identification of six modes is possible when $k_x\to
-\infty$, too, thus defining six independent functions $\cx_n(k_x)$,
solution of (\ref{systdiff2}).\\
The two bases of solutions, ($\cx^m$) and ($\cx_n$), satisfy the same
differential system and are consequently linked with a matrix of complex
constants ${\cal C}_{mn}$ defined as
\begin{equation}
\cx^m(k_x) =\sum {\cal C}_{mn}\cx_n(k_x). \label{coupling}
\end{equation}
The coupling constants satisfy some symmetry relations, taking advantage
of the fact that if $(\cx_x,\cx_y,\cx_z)(k_x)$ is a solution, then
$(\cx_x,-\cx_y,-\cx_z)(-k_x)$ and  $(\cx_x^*,\cx_y^*,-\cx_z^*)(k_x)$ are
solutions too.\\
The coupling constants depend on the parameters
($k_y,\alpha,\beta,\Omega,A$), and describe how a perturbation with a
definite initial polarization ($K_x(t=0)=k_x\to-\infty$) is finally
projected, because of differential rotation, onto the six different
polarizations when $K_x(t)\to+\infty$.\\
The so-called ``asymptotic states'' are relevant as soon as
$|K_x(t)|>k_{\infty}$. A crude majoration
($\bk_{\infty}<1+\tilde\Omega$)  shows that the formulation in terms of
coupling constants is convenient as long as we consider time-scales
longer than $|1/A|$. They can account for the following features~:
\par (i) The situation where the initial perturbation conserves its
identity during evolution (referred as ``adiabatic'' in
Fig.~\ref{fclassicadiabatic} and Fig.~\ref{ftransientadiabatic}) is
represented with a diagonal  $({\cal C}_{mn})$ matrix. The WKB method
allows a detailed description of this situation in Sect.~8.3 in the case
of weak differential rotation. The self coupling constant goes to zero
like ${\cal C}_{\rm PP}(A)\sim\exp(-|{\rm cte}/A|)$ when $A\to 0$
because the growth rate of the instability at $|K_x|<k_\infty$ is lower
than $\omega_\Delta$ during a time $\sim k_\infty/(Ak_y)$ (see
Fig.~\ref{fcppA}).
\par (ii) The linear coupling of modes corresponds to a non-diagonal
$({\cal C}_{mn})$ matrix (Sect.~8.4). For example, ${\cal C}_{PS}$ and
${\cal C}_{PA}$ respectively indicate the amount of magnetosonic and
stable  Alfvenic waves generated by the shearing of Parker instability,
as illustrated in Fig.~\ref{fwave} and Fig.~\ref{fturnover}.
\par (iii) The turnover occurs when the coupling constant ${\cal C}_{\rm
PP}$ relative to the Parker unstable mode is negative (see
Fig.~\ref{fturnover} and Sect.~8.5).
\par (iv) A transient amplification due to differential rotation would
correspond to the presence of some remarkably high
 coupling constants (${\cal C}_{PP}<-1$ in the case of
overamplification, $|{\cal C}_{AA}|>1$ in Fig. \ref{amplif} of
Sect.~9).\\
\begin{figure}
\picture 87.8mm by 63.8mm (fcppA)
\caption[]{Dependence of the Parker self-coupling constant ${\cal
C}_{PP}$ on the shear parameter $-A/\Omega$, calculated numerically with
$\alpha=1$, and $\beta=0$ (full line) and $\beta=1$ (dotted line). The
turn-over occurs in both cases for $-A/\Omega>0.42$}  \label{fcppA}
\end{figure}

\subsection{Analysis of the adiabatic evolution~: WKB approximation}
If the shearing motion is slow enough that a perturbation conserves its
{\it identity} throughout its sheared evolution (\ie if the non-diagonal
coupling constants are vanishing), we can obtain a dispersion relation
describing its instantaneous frequency. In this case, not only is the
{\it identity} of the perturbation asymptotically well defined, but the
WKB approximation allows us to ``follow'' its evolution analytically, at
any time. The WKB formalism is presented in Sect.~8.3.1, and applied to
the weak shear limit in Sect.~8.3.2., confirming the study of Shu
(1974). A more surprising result is presented in Sect.~8.3.3, where the
transient stabilizing effect of the differential force is strong
although the differential rotation is weak.

\subsubsection{WKB formalism~: from poles to saddle points}
The homogeneous differential equation associated to
Eq.~(\ref{Frobenius}) may be written in its canonical form~:
\begin{equation}
\left\{{\p^2\over\p\tom^2}+W(\tom^2)\right\}(f^{1/2}\bx_H)=0,
\label{canon}
\end{equation}
where $W(\tom^2)$ is defined as
\begin{equation}
W(\tom^2)\equiv g(\tom^2)-{1\over2}{\p^2\log f(\tom^2)\over\p\tom^2}-
{1\over4}\left({\p\log f(\tom^2)\over\p\tom}\right)^2.\nonumber
\end{equation}
Its solutions can be approximated, using the WKB method, with a linear
combination of the two independent functions~:
\begin{equation}
\bx_\pm(\tom)\equiv {1\over (f^2W)^{1/4}}\exp\left(\pm i\int^{\tom}
W^{1/2}(\tom'^2)\d\tom'\right).\label{WKB}
\end{equation}
Using this approximation, the inverse Laplace transform
(\ref{invLaplace}) becomes~:
\begin{eqnarray}
\tx(k_x,t)={1\over2\pi}\int_{ip-\infty}^{ip+\infty}
\left(\mu_+\e^{i\Psi_+}- \mu_-\e^{i\Psi_-}\right)\d \omega,
\label{saddle}
\end{eqnarray}
where the phase $\Psi_\pm(\tom)$ is defined in a dimensionless form as
\begin{equation}
\Psi_\pm(\tom,t)\equiv
-{\tilde\Omega\over\bk_A^2\bk_y}(\bk_x-2\bk_yAt)\tom  \pm\int^{\tom}
W^{1/2}(\tom'^2)\d\tom'.\nonumber
\end{equation}
If satisfied, the WKB criterion defined below ensures that the functions
$\mu_\pm(\tom)$ vary slowly. The major contributions to the integral
(\ref{saddle}) are thus given by the saddle points $\tom_\pm(t)$ given
by~:
\begin{equation}
{\p\Psi_\pm\over\p\tom}(\tom^2_\pm(t))=0,\nonumber
\end{equation}
which is equivalent to~:
\begin{equation}
{\bk_A^2\bk_y\over\tilde\Omega}W^{1/2}(\tom^2_\pm(t))=\pm
(\bk_x-2\bk_yAt). \label{WKBdisp2}
\end{equation}
The integration contour may be moved in order to pass through these
saddle points, and follow the steepest descent path between them. This
formulation allows us to interpret Eq.~(\ref{WKBdisp2}) as a dispersion
relation whose roots are time-dependent. These saddle points are the
relevant {\it instantaneous} growth rates or frequencies which used to
be the poles of the Laplace transform $\bx(\tom)$ in the absence of
differential rotation.\\
We show in Appendix~D that the WKB-approximation is relevant in two
different cases~:
\begin{equation}
{\rm ~~~(i)~~~when }\;\;
|\tom|\gg\left({3\bk_A^4\over8\tilde\Omega^2\bk_y^2}
{1+\alpha\over\alpha}\right)^{1/4},\label{WKBvalid0}
\end{equation}
the asymptotic behaviour of $W(\tom^2)$, together with
Eq.~(\ref{WKBdisp2}), gives the two magnetosonic saddle points
($\tom_{\rm S}(t)\sim (1+(2\alpha)^{-1})^{1/2}\bar K_x(t)$ and its
opposite), producing two contributions to the inverse Laplace transform,
noted $\tx_{\rm sad.} (k_x,t)$ in Sect.~8.2~:
\begin{eqnarray}
\tx_{\rm sad.} (k_x,t)&\sim&\mu_{+}(\tom_{\rm
S}(t))\e^{i\Psi_{+}(\tom_{\rm S}(t))}\\ &+&\mu_{-}(-\tom_{\rm
S}(t))\e^{i\Psi_{-}(-\tom_{\rm S}(t))},
\end{eqnarray}
where
\begin{eqnarray}
\Psi_{\pm}(\tom_{\rm
S}(t))\sim\mp{\tilde\Omega\over\bk_A^2\bk_y}\left(1+2\alpha\over2\alpha\right)^{1/2}
{\bar K_x^2(t)\over2} {\rm~~~as~~} |K_x|\to\infty.\nonumber
\end{eqnarray}
When performing the Fourier transform $k_x\to x$, the
contributions of the magnetosonic $\tom$-saddle points produce four
$k_x$-saddle points in the Fourier integral (\ref{Fourierinv}),
corresponding to four magnetosonic waves propagating radially in both
directions, on both sides of corotation, with the magnetosonic speed
$(a^2+V_{\rm A}^2)^{1/2}$.\\
\par (ii) The WKB approximation is correct for bounded values of $\tom$
(but away from the singular points), only if the strength of
differential rotation is limited with~:
\begin{eqnarray}
{\bk_A^2\over\tilde\Omega}\ll1 \Longleftrightarrow T_{\rm Parker}\ll
T_{\rm Shear}. \label{WKBvalid}
\end{eqnarray}
This WKB criterion can be understood as a measure of the time-scale over
which the polarizations of the eigenvectors change ($T_{\rm Shear}$),
compared to the intrinsic time-scale of the solution ($T_{\rm
Parker}$).\\

\subsubsection{Classical adiabatic evolution}
The simplest case satisfying the WKB criterion occurs when the
centrifugal force is negligible compared to the magnetic radial tension
\begin{equation}
\bk_A^2\ll1. \nonumber
\end{equation}
As in the study of Shu (1974), the shear parameter  disappears from the
instantaneous dispersion relation (\ref{WKBdisp2}), except in the
implicit time-dependence of $K_x(t)$~:
\begin{equation}
Q_3(\tom^2)=\bar K_x^2(t)\Delta(\tom^2). \nonumber
\end{equation}
It identifies with the dispersion relation of a disk with uniform
rotation, where $\bar K_x(t)$ replaces $\bk_x$. The results of Sect.~5
are consequently relevant in this approximation, and the numerical
simulation in Fig.~\ref{fclassicadiabatic} is a typical illustration of
this behaviour.\\
This kind of evolution is the most natural to be expected from the time
dependence of the radial wavenumber; however it is relevant only for
unrealistic low shear; for example we find that, if $\alpha=1$, this
simple evolution occurs only for $0<-A/\Omega\le 0.02$).

\subsubsection{Analytical evidence for the transient shearing
stabilization} As mentioned in Sect.~7.2, $\alpha\ll1$ allows the
differential force to remain comparable to the magnetic tension
($\bk_A\sim \bk_y$), while the Parker time still remains small compared
to the shear time to ensure an adiabatic evolution ($1/\tilde\Omega\sim
\alpha^{1/2}\ll1$)~:
\begin{equation}
T_{\Omega}\ll T_{\rm Parker}\ll T_{\rm Shear}. \nonumber
\end{equation}
As discussed in Appendix D, the equations obtained for the   components
of $\bx$ are equivalent; however the WKB method introduces  artificial
singularities in the equations for all but $\bx_x$; thus we have chosen
to work on the latter equation. \\
In the limit of $\tilde\Omega\gg1$, and for bounded values of $\bar
K_x(t)$, we can simplify the dispersion relation (\ref{WKBdisp2})
as~:
\begin{eqnarray}
2\alpha\tom^6&-&\left(2\alpha\tilde\Omega^2+\bar
K_x^2+\bk_y^2+{1\over4}\right)\tom^4
+{\tilde\Omega^2\over4}\tom^2-\bk_y^2R(\bar K_x^2)=0,\nonumber
\end{eqnarray}
with
\begin{equation}
R(\bar K_x^2)=\bk_y^4-(\bk_{\rm Q}^2-\bar K_x^2)\bk_y^2 -\bar
K_x^2\bk_{\rm P}^2+{\bk_A^2\over4}. \nonumber
\end{equation}
In the limit $\tilde\Omega\gg1$, the roots of this instantaneous
dispersion equation can be approximated as:
\begin{eqnarray}
\tom_{\rm S}^2,\tom_{\rm A}^2&\sim&\tilde\Omega^2\gg1\nonumber\\
\tom_{\rm P}^2&\sim&{4\bk_y^2R(\bar K_x^2)\over\tilde\Omega^2}\ll1
\label{wstable}
\end{eqnarray}
The latter root identifies with the unstable Parker mode ($R<0$) when
$\bk_A=0$ and $\bk_y<\bk_{\rm Q}$.\\ The differential rotation is
clearly stabilizing ($R>0$) at low $\bar K_x(t)$ as soon as
\begin{equation}
\bk_A^2> 4\bk_y^2(\bk_{\rm Q}^2-\bk_y^2). \label{shearmini}
\end{equation}
In particular, all the azimuthal wavelengths are stabilized at $\bar
K_x(t)=0$ if $\bk_A>k_{\rm Q}^2$.\\ We can draw a parallel between this
criterion for transient stability and the instability-criterion found by
Balbus \& Hawley (1991) for accretion disks with a vertical magnetic
field, to outline the decisive role played by the configuration of the
equilibrium magnetic field. We shall come back to this point in
Sect.~9.\\
The stabilization lasts while $|\bar K_x(t)|<\bar K_o$ (see
Fig.~\ref{fWwk}), with
\begin{eqnarray}
\bar K_o^2={\bk_A^2/4 -\bk_y^2(\bk_{\rm
Q}^2-\bk_y^2)\over \bk_P^2-\bk_y^2}.\label{klimite}
\end{eqnarray}
\begin{figure}
\picture 87.8mm by 63.8mm (fWwk)
\caption[]{The Parker frequency $\omega_{\rm P}$ is always imaginary in
the classical case (dotted line). It becomes real for $|K_x(t)|<K_0$
(full line) if the differential force is strong compared to the radial
magnetic tension, corresponding to a transient stabilization of the
instability. The Parker instability always dominates asymptotically
(dashed line)}  \label{fWwk}
\end{figure}
We can estimate the overall phase change $\varphi$ during this
stabilizing oscillatory period ($t_1\le t\le t_2$ while $\tom_{\rm
P}^2(t)\ge 0$)~:
\begin{eqnarray}
\varphi\equiv\int_{t_1}^{t_2}\omega_3(t)\d t\nonumber
\end{eqnarray}
In our dimensionless variables, $\varphi$ becomes~:
\begin{eqnarray}
\varphi={\tilde\Omega\over\bk_A^2\bk_y}\int_{-\bar K_o}^{+\bar K_o}
\tom_{\rm P}(\bar K_x)\d \bar K_x.\nonumber
\end{eqnarray}
Using the estimate (\ref{wstable}) for $\tom_{\rm P}(\bar K_x)$, we find
that the phase change is independent of the small  parameter
($1/\tilde\Omega$) of the WKB approximation:
\begin{eqnarray}
\varphi=\left\{1+{4\bk_y^2(\bk_{\rm Q}^2-\bk_y^2)\over\bk_A^2}\right\}
\varphi_{lim}, \label{dephasage}
\end{eqnarray}
with
\begin{equation}
\varphi_{lim}= {1\over (\bk_P^2-\bk_y^2)^{1/2}}{\pi\over4}. \nonumber
\end{equation}
In particular, $\varphi_{lim}=\pi/2$ when $\bk_y=\bk_{\rm Q}$.\\ If the
optimum value of the azimuthal wavenumber $\bk_y=\bk_\Delta$ is chosen,
we find
\begin{equation}
\varphi_{lim}={\pi\over\{(1+\beta/\alpha)(5+\beta/\alpha)\}^{1/2}}
<{\pi\over\surd5}. \nonumber
\end{equation}
We must distinguish here between two different ranges of azimuthal
wavenumbers~:
\par (i) $\bk_{\rm Q}<\bk_y<\bk_P$ : we know from Sect.~3-5 that in this
case, the Parker instability becomes stable at low $\bk_x$, even without
differential rotation. Consequently, the longer the time spent at low
$\bar K_x$, i.e. the lower the shear parameter $k_A$, the larger the
number of oscillations $[\varphi/2\pi]$. In this context, a turn-over
occurs (${\cal C}_{\rm PP}<0$) only if $[\varphi/\pi]$ is odd. The
minimum phase change is $\varphi_{lim}>\pi/2$, so there is always at
least one oscillation.
\par  (ii) $\bk_y<\bk_{\rm Q}<\bk_P$ : the differential force is
responsible for a stabilization of the Parker instability at low $\bar
K_x$ if the shear is high enough (see Eq.~(\ref{shearmini})). The higher
the shear parameter $\bk_A$, the shorter the time spent at $|\bar
K_x|<\bar K_o$, but the higher the oscillation frequency $\tom_{\rm P}$
and the higher $\bar K_o$. All together, these opposite effects result
in a phase change $\varphi$ which increases with the shear parameter
$\bk_A$, but is bounded by the maximum value $\varphi_{lim}<\pi/2$, thus
implying that no turn-over can occur.\\
The asymptotic Parker instability favours the azimuthal wavenumber
$\bk_y=\bk_\Delta<\bk_{\rm Q}$ given in Appendix~A, so that no turn-over
can be expected in that WKB limit. Nevertheless, a transient
stabilization occurs, whose duration is approximately~:
\begin{equation}
\tau\sim \left|{\Omega\over A}\right|^{1/2}{H\over
V_{\rm A}}\sim \bk_A T_{\rm Shear}\gg T_{\rm Parker}\gg T_{\Omega}.
\nonumber
\end{equation}
This transient adiabatic stabilization by the differential force was
verified in numerical simulations in the  extreme mathematical limit
($\alpha\ll1$); it was also found to be qualitatively correct for
$\alpha=$ 0 to 5, as long as  $T_{\rm Parker}/ T_{\rm Shear} \le 0.2$
(see Fig.~\ref{ftransientadiabatic}).

\subsection{Non-adiabatic evolution: linear coupling}
If $T_{\rm Parker}\sim T_{\rm Shear}$, the WKB approximation fails. A
solution which is initially polarized along the asymptotic Parker
eigenvector cannot ``follow'' this fast time-varying eigenvector during
the period of transition $|\bar K_x|<\bk_\infty$. The solution
consequently ``looses'' its leading polarization and projects along all
other eigenvectors at $\bar K_x>\bk_\infty$. This explains
mathematically the generation of magnetosonic as well as Alfven waves,
as a result of this projection.\\
This coupling may be understood within the description of the modes in
terms of saddle points, in the following manner~:\\
Considering the {\it distinct} instantaneous growth rates and
frequencies given by the saddle points Eq.~(\ref{WKBdisp2}) supposes
that the typical lengthscale of the saddle $\delta\tom$, \ie the
distance over which the integrand of the  inverse Laplace transform
decreases away from the saddle point, is small  compared to the
distances to the other saddle points. The saddle is described by an
expansion in the vicinity of the saddle point $\tom_\pm(t)$~:
\begin{equation}
\Psi(\tom,t)=\Psi(\tom_\pm(t))+\left({\tom-\tom_\pm(t)\over
\delta\tom}\right)^2+O\{(\tom-\tom_\pm(t))^3\}, \nonumber
\end{equation}
with
\begin{equation}
\delta\tom^2={2\bk_y\over\bar K_x\tom_\pm(t)}\left({\p\log
W\over\p\tom^2}\right)^{-1}{\bk_A^2\over\tilde\Omega}. \nonumber
\end{equation}
Consequently, the saddle points are well isolated as long as the
WKB-criterion $\bk_A^2/\tilde\Omega\ll1$ is satisfied. Moreover, the
higher the radial wavenumber $|\bar K_x(t)|$, the narrower the  width of
the saddle. Increasing $|\bar K_x(t)|$ makes the two saddle points
$\tom_{\rm A},\tom_{\rm P}$ converge to their asymptotic value
$\tod,\todp$, where the WKB approximation breaks down, because of the
singularity.\\
When $\bk_A^2/\tilde\Omega\sim1$, the width of the saddles increases so
much at low $|\bar K_x(t)|$, that it becomes impossible to {\it
identify}, in the $\tom$-complex plane, well-isolated contributions to
the inverse Laplace integral~: the three modes cease to be
distinguishable at low $|\bar K_x(t)|$.\\
The linear coupling between the modes is a consequence of the high rate
of shear, rather than a consequence of the mechanism of the Parker
instability~: this coupling is still present in the stable configuration
($k_y>k_P$), for initial perturbations polarized along either the stable
Parker, or Alfvenic, or magnetosonic mode (see Fig.~\ref{fcoupling}).\\
It may occur in any problem where shearing is present, as soon as the
rate of shear is high enough (see Ryu \& Goodman 1992).\\
\begin{figure}
\picture 87.8mm by 63.8mm (fcoupling)
\caption[]{Linear coupling in a stable configuration ($k_y>k_{\rm P}$)~:
the initial magnetosonic wave (fast radial oscillations) couples at
$|\bar K_x(t)|<1$ to the Alfven mode (${\cal C}_{\rm SA}\ne 0$), and to
the stable Parker mode (${\cal C}_{\rm SP}\ne 0$)}  \label{fcoupling}
\end{figure}

\subsection{analytical formulation of the turn-over}
The calculation of Sect.~8.2 allows us to give an analytical formulation
of the turn-over phenomenon~: a perturbation with a definite radial
wavenumber $k_{0x}$, initially polarized along Parker asymptotic
eigenector $\xi_{\rm P}(k_x)=[\xi_{{\rm P}_x},\xi_{{\rm P}_y},\xi_{{\rm
P}_z}]$ (given in Sect.~5 and normalized with $\xi_{{\rm P}_x}\sim
1/\bk_x$), grows like
\begin{equation}
\tx(t,x=0)\sim \xi_{\rm P}(k_{0x}-2Ak_yt)\e^{-i\od
t},\nonumber
\end{equation}
while $k_{0x}-2Ak_yt\ll-\bk_\infty$. We know from Sect.~8.2.1 that it
will finally grow like
\begin{eqnarray}
\tx(t,x=0)&\sim& \e^{-i\od (t-{k_{0x}\over2Ak_y})} \left|
\begin{array}{c} \Lambda_x^{\rm P}(k_{0x})(H/V_{\rm
A}t)+O(1/t^2)\nonumber\\
\Lambda_y^{\rm P}(k_{0x})+O(1/t)\\
\Lambda_z^{\rm P}(k_{0x})+O(1/t)
\end{array}
\right.\\
&\sim& {\cal C}_{\rm PP}\xi_{\rm P}(k_{0x}-2Ak_yt)\e^{-i\od t}\nonumber
\end{eqnarray}
so that we can calculate the self coupling constant as
\begin{equation}
{\cal C}_{\rm PP} =\lim_{k_{0x}\to-\infty}{2\bk_y AH\over V_A}
e^{-i{\tom\tilde\Omega\bk_{0x}\over\bk_y\bk_A^2}} \Lambda^{\rm
P}_x(k_{0x}).
\end{equation}
${\cal C}_{\rm PP}$ is a measure of the damping effect of the
oscillation phase. It is a function of $\alpha,\beta,\bk_y,\tilde\Omega$
and $\bk_A$.\\
${\cal C}_{\rm PP}$ would be equal to $1$ if differential rotation had a
negligible effect on the Parker growth. In fact, it can be much smaller
than $1$, and even negative when the turn-over occurs, as was seen in
Sect.~7.4.\\
${\cal C}_{\rm PP}$ vanishes in the intermediate case, when the
differential force modifies the polarization of the solution so that its
final projection on the Parker mode vanishes. Beyond  the parameter range
where it vanishes, ${\cal C}_{\rm PP}$ becomes negative: the  coupling
with the waves reverses the direction of the Parker mode. It  is
remarkable that, in that case, $|{\cal C}_{\rm PP}|$ can become larger
than one,  so that the coupling with the waves {\it increases} the
strength of  the Parker instability. We will refer to this case as {\it
overamplified  turnover} (see Fig.~\ref{fPhi_A}).\\
\begin{figure}
\picture 87.8mm by 63.8mm (fPhi_A)
\caption[]{$\alpha$-dependence of ${\cal C}_{\rm PP}$, for a Keplerian
rotation ($-A/\Omega=0.75$ in full line), and a ``flat'' rotation curve
($-A/\Omega=0.5$ in dashed line). A strong magnetic field impedes the
turnover process. The overamplification occurs at low magnetic pressure
only if the differential rotation is high enough}  \label{fPhi_A}
\end{figure}
We show in Appendix~C that the analytical expression of ${\cal C}_{\rm
PP}$ depends only on the behaviour of one single fractional function
$W(\tom^2)$ on the semi-axis ranging from $\tod-\infty$ to $\tod$~:\\
Let $\zeta$ be the unique solution of the canonical expression
(\ref{canon}) of the homogeneous differential equation
\begin{equation}
\left\{{\p^2\over\p\tom^2}+W(\tom^2)\right\}\zeta=0,
\end{equation}
which vanishes at $\tom\to \tod-\infty$. Its behaviour near $\tod$
defines two complex numbers ($\lambda,\mu$) so that~:
\begin{equation}
\zeta=(\tom-\tod)^{1/2}\left\{\mu\log\left({\tom-\tod\over i}\right)
+\lambda + O(\tom-\tod)\right\}. \nonumber
\end{equation}
We show in Appendix~C that
\begin{equation}
{\cal C}_{\rm PP}= P_x(\tod)
\left[{\p\Delta\over\p\tom}(\tod)\right]^{-1}\left(1- {\pi\over {\rm
Im}\left({\lambda\over\mu}\right)}\right), \label{analyticturn}
\end{equation}
where $P_x(\tod)$ and $\Delta$ do not depend on the strength of
differential rotation. The turn-over limit is
\begin{equation}
{\rm Im}\left({\lambda\over\mu}\right)= \pi.\label{turnover}
\end{equation}
In particular, one can verify that in the limit of vanishing shear, the
converging solution $\zeta$ is well approximated, in the vicinity of
$\tod$, by a Bessel function
\begin{equation}
\zeta(\tom)\sim(\tom-\tod)^{1/2}H_0^{(2)}(\nu(\tom-\tod)^{1/2}),
\end{equation}
where  $\nu$ is defined as
\begin{equation}
W(\tom^2)={1\over4}\left({1\over (\tom-\tod)^2}+{\nu^2\over \tom-\tod}
+O(\tom-\tod)\right).
\end{equation}
The formula (\ref{analyticturn}) then ensures that ${\cal C}_{\rm PP}\to
0$ in the limit of vanishing shear.\\
This mathematical formulation allows an easy numerical study of the
dependence of the criterion on the various parameters, especially
$\alpha,\beta,\tilde\Omega$ and $\bk_A$.\\
Furthermore, it shows the integral nature of the phenomenon, which can
be seen as the requirement that the stabilizing oscillation at low $\bar
K_x(t)$, outlined in Sect.~8.3.3 has made the polarization rotate so as
to be in exact opposite phase ($\varphi=\pi/2$) with the Parker unstable
polarization.\\

\section{Simplified modelization applied to radially localized
perturbations }
{}From the first of our mathematical descriptions (\ref{initialshape}), we
see  that, taking the arbitrary functions $P_j(k_x)$ of the form
$\delta(k_x-k_0)$, the  solutions $\check\xi(k_0-2Ak_yt)$ described in
the previous sections directly give  the time evolution of a
perturbation varying initially as  exp$(ik_0x)$. This is of course not a
very realistic initial condition  for the Parker instability. It would
more likely start from a localized  perturbation as can occur in the
galactic disk, such as the bubble from a  supernova explosion, or the
evolution of a molecular cloud. An initial  perturbation that is
localised in $x$, say $\delta(x)$ or a narrow gaussian,  corresponds to
a Fourier transform $\hat\xi(k_x,t=0)$ which is a constant or a wide
gaussian. The amplitude of such a perturbation is initially the same at
all the radial wavelengths, and its polarization projects along the six
modes of evolution. In order to describe this time evolution, we use the
fact that differential rotation  modifies the initial perturbation most
effectively in the limited time range such  that
$|K_x(t)|<1+\tilde\Omega$, and we modelize its effect by a jump at
$k_x=0$; thus we neglect the details of the coupling with the waves,
and  retain only the effect of the value of ${\cal C}_{\rm PP}$, \ie
decrease of the  strength of the instability, turnover, or overamplified
turnover. \\
At a first approximation, we simplify the behaviour described before as
follows~:
\par (i) the transition at $|HK_x|\ll \bk_\infty$ is supposed to be
instantaneous.
\par (ii) the polarization of the Parker eigenvector is
independent of $k_x$ and contained in the $(y,z)$ plane.
\par (iii) a perturbation of initial amplitude $\xi_o$ with $k_x>0$
grows like~:
\begin{eqnarray}
\xi(t)=\xi_o \e^{(k_x-2Ak_yt)x}\e^{-i\omega_\Delta t} \mbox{ for all }
t>0. \nonumber
\end{eqnarray}
\par (iv) a perturbation of initial amplitude $\xi_o$ with $k_x<0$ grows
like~:
\begin{eqnarray}
\xi(t)&=&\xi_o \e^{(k_x-2Ak_yt)x}\e^{-i\omega_\Delta t} \mbox{ while }
0<t<{k_x\over 2Ak_y}\\ \xi(t)&=&{\cal C}_{\rm PP}\xi_o
\e^{(k_x-2Ak_yt)x}\e^{-i\omega_\Delta t} \mbox{ for } t>{k_x\over 2Ak_y}
\end{eqnarray}
Simplifying the $k_x$-dependence at the scale $\Delta
k<(1+\tilde\Omega)/H$ implies a loss of information about any occurrence
at spatial scales $\Delta x >H/(1+\tilde\Omega)$.\\
Any initial perturbation can be written~:
\begin{equation}
\xi(t=0,x)=\int_{-\infty}^{+\infty}\tx_o(k_x)\e^{ik_xx}\d k_x.\nonumber
\end{equation}
Within our approximation, we find~:
\begin{eqnarray}
\xi(t,x) &=&\e^{-i\od t}\e^{-2iAk_ytx}
\left\{\xi(0,x)\tvi(12,12)\right.\nonumber\\
&+&\left.( {\cal C}_{\rm PP}-1)\int_{2Ak_yt}^0 \xi_o(k_x)\e^{ik_xx}\d
k_x\right\}. \nonumber
\end{eqnarray}
The case of a radially localized perturbation, corresponding to
$\tx_o(k_x)\sim $ const., leads to
\begin{eqnarray}
\xi(t,x)=\e^{-i\omega_\Delta t-2iAk_ytx}\left\{\xi(0,x)+ {1-{\cal
C}_{\rm PP}\over ix}(\e^{2iAk_ytx}-1)\right\}. \nonumber
\end{eqnarray}
The asymptotic behaviour of such a perturbation is~:
\begin{equation}
\xi(t,x)\sim {\cal C}_{\rm PP} \e^{-i\omega_\Delta
t}\delta(x=0)\;\;\mbox{ when }\;t\to+\infty,\nonumber
\end{equation}
so that a turn-over occurs even though all the radial wavelengths are
superimposed. The sign of ${\cal C}_{\rm PP}$ consequently rules the
general direction of the perturbed gas, and ${\cal C}_{\rm PP}\not= 1$
leads to high contrasts of density on very short radial scales (see
Fig.~\ref{fdensit}).\\
\begin{figure}
\picture 87.8mm by 63.8mm (fdensit)
\caption[]{Qualitative description of the turn-over of a perturbation
which is initially localized at x=0, in our simplified model. As time
grows, the initial ``dirac'' profile (dotted line) is widened by the
effect of differential rotation (thin full line), and finally becomes
inverted (thick full line)} \label{fdensit}
\end{figure}

\section{Comment on the instability found by Balbus \& Hawley (1991)}
We have noted in Sect.~8.3.3 that the same criterion found by
Chandrasekhar (1960) and Balbus \& Hawley (1991) for the instability of
disks with a vertical magnetic field appears here as a criterion for
transient stabilization. Our calculations appear in fact  more related
to that of Tagger \etal (1992a, 1992b) for non-axisymmetric
instability, since in both cases we are dealing with transient
phenomena.\\
This criterion can be written in physical variables as:
\begin{equation}
4A\Omega+k_y^2V_{\rm A}^2<0,\label{BHawley}
\end{equation}
which can be compared (at small $k_x$) to the instability criterion of
Chandrasekhar (1960), used by Balbus and Hawley:
\begin{equation}
4A\Omega+k_z^2V_{\rm A}^2<0.\label{BHawley}
\end{equation}
In both cases the magnetic term, due to the energetic cost of bending
field lines, depends on the wavenumber parallel to the equilibrium
field. This transient stabilization was discussed analytically in
Sect.~8.3.3 in the limit of an adiabatic evolution only, i.e. when the
shear time is long enough compared to the Parker time.\\
Nevertheless, we must still consider a possible destabilizing effect of
strong (realistic) differential rotation in cases which are not
dominated by the  Parker instability; this can be tested numerically for
stable values of the azimuthal wavenumber ($k_y>k_{\rm P}$),  \ie for
small enough wavelengths. It allows us to dissociate the growth due to
the Parker instability from the possible transient shearing growth. \\
In this range of azimuthal wavenumbers, the three modes of oscillation
are a magnetosonic mode, and two purely magnetic modes, the slowest
being the formerly unstable Parker mode.\\
In order to test the criterion (\ref{BHawley}), and to avoid the natural
stabilizing influence of a strong magnetic tension, we consider weak
magnetic fields~:
\begin{equation}
V_{\rm A}^2<\left|{4A\Omega\over k_{\rm P}^2}\right|. \nonumber
\end{equation}
The analytical study performed in Sect.~8.3.3 is still valid if
$k_y>k_{\rm P}$ and in that case, differential rotation appears to
remain stabilizing. \\
A transient amplification of the Alfvenic and Parker modes is observed
in the numerical simulations, when the shear time is short compared to
the Alfven time across the disk height, ensuring a non-adiabatic
behaviour. The change of polarization between the asymptotic states
($|K_x|(t)<k_{\infty}$) is so fast that the initial magnetic wave reacts
to it as if it were a shock. This brief impulsion is essentially
produced by the differential force (strong compared to the weak magnetic
tension) in the radial direction, giving birth to a high azimuthal speed
because of the Coriolis force (see Fig.~\ref{amplif}). \\
This dominant and amplifying effect of the differential force at low
$|K_x(t)|$ appeared also in the case of the unstable Parker mode, as a
source of overamplification.
\begin{figure}
\picture 87.8mm by 63.8mm (amplif)
\picture 87.8mm by 63.8mm (amplifv)
\caption[] {Amplification of an Alfven mode in a non-adiabatic regime at
low magnetic pressure. With $\alpha=0.0025, \beta=0, -A/\Omega=0.5$, we
get the amplification ${\cal{C}}_{\rm AA}\sim 10$ on the displacements
(above) as well as on the speeds (below). The amplification process is
clearly taking place in the ($x,y$) plane, under the control of the
strong inertial forces (both differential and Coriolis)}  \label{amplif}
\end{figure}
It is interesting to note that differential rotation, combined to MHD
behaviour may lead to opposite effects, according to the strength of the
magnetic field~:
\par (i) differential rotation transiently stabilizes the Parker
instability if the shear time is much longer than the Alfven time.
\par (ii) it may destabilize the Parker and the Alfven modes if the
shear time is much shorter than the Alfven time, acting as a short
impulsion on the radial displacement.\\
We also note that these two cases of magnetic field geometry, \ie purely
toroidal (azimuthal) or purely poloidal (vertical) are of course extreme
limits and cannot claim to be very realistic configurations. On the
other hand, one easily finds that the constraints of ideal MHD
equilibrium for a disk with both toroidal and poloidal fields require
that the former be independent of $z$. Instability criteria in this
configuration for axisymmetric modes (the simplest case) turn out to be
more complicated (Dubrulle \& Knobloch 1993). Since these equilibria do
not have the stratified character necessary for the Parker instability
to occur, it is unlikely that a more detailed theory of the connection
between these two types of instability can be elaborated on.

\section{The range of galactic parameters}
Although several types of evolution may take place in a
sheared disk embedded in an azimuthal magnetic field (adiabatic,
transient adiabatic stabilization, linear coupling, turnover,
overamplification), the range of galactic parameters allowed by
observations reduces the number of possibilities.
\par (i) An adiabatic evolution is ruled out in our region of the galaxy
because of the high shear ($-A/\Omega\sim .5-.75$).
\par(ii) The high magnetic pressure observed in our galaxy
($\alpha>0.5$) rules out the possibility of an overamplification (see
Fig.~\ref{fPhi_A}).
\par (iii) We must take into account the gravitationnal influence of the
stars of the midplane disk, in diminishing the scale height of the gas
(see Sect.~5, $\tilde\Omega\sim 0.7$ according to observations). Figure
\ref{fhauteur} shows the high dependence of the coupling constant ${\cal
C}_{\rm PP}$ on the parameter $\tilde\Omega$, and  precludes the
possibility of a turn-over with the galactic values of the parameters.
Moreover, it shows the influence of the cosmic rays in destabilizing the
Parker mode, hence favouring  positive values of ${\cal C}_{\rm PP}$,
lowering the transient stabilization by the Coriolis force and
differential rotation, and impeding the turn-over.\\
In the frame of this simple model, the coupling to Alfvenic and
magnetosonic waves seems to be the most realistic picture of evolution
of the Parker instability.\\
On the other hand, as mentionned in Sect.~2.1, the observed scale height
of the magnetic field is much larger than in our model. This might
strongly affect the limits of the parameter ranges defined here, but its
effect cannot be estimated within the present model and should be
obtained from numerical simulations.\\
\begin{figure}
\picture 87.8mm by 63.8mm (fhauteur)
\caption[]{Taking into account the gravity of the midplane stars
diminishes the parameter $\tilde\Omega=2\Omega H/V_{\rm A}$ which is
proportional to the scale height of the gas. The evolution becomes
adiabatic in the limit of an infinitely thin disk of
gas($\tilde\Omega\to0$). With a realistic scale height
($\tilde\Omega\sim 0.7$), the evolution is still coupled to
magnetosonic and Alfven waves, but no turn-over occurs. These
simulations correspond to $\alpha=1$, with $-A/\Omega=0.5$ (thin lines),
$-A/\Omega=0.75$ (thick lines), $\beta=0$ (dotted lines) and $\beta=1$
(full lines)} \label{fhauteur}
\end{figure}

\section{Discussion and conclusion}
We have been able in this paper to clarify both the mathematical
structure and the physical properties of perturbations evolving in a
differentially rotating disk embedded in an azimuthal magnetic field.
Although some of our results simply confirm, with a different approach,
the analysis performed by Shu (1974), we have pointed out some new
results which we summarize here.
\par (i) The interpretation of differential rotation in term of a
differential force and a time-dependent radial wavenumber allows a more
physical understanding of the mathematical calculations.
\par (ii) The coupling of modes introduced by differential rotation and
mentionned by Shu (1974) is here given an adequate formalism to allow
physical investigations such as the quantitative amount of magnetosonic
waves generated by the Parker instability, and conversely.
\par (iii) A WKB approach, in the case of both low shear and low
magnetic field, leads to a dispersion equation where the transient
stabilizing effect of the differential force appears analytically. In
the case of low shear, the contribution of each mode is mathematically
identified in the complex plane with a saddle point contribution. The
asymptotic behaviour is defined by the contributions of simple poles at
the Alfvenic and Parker frequencies, in addition to the magnetosonic
saddle point diverging to real infinity.
\par (iv) The mathematical structure of the problem is reduced to a
second order differential equation. In particular, the self-coupling
constant ${\cal C}_{\rm PP}$ of the Parker mode is derived analytically,
and its simplest formulation is obtained.
\par (v) A numerical study of the evolution of perturbations leads us to
define several qualitative behaviours : adiabatic evolution with a
transient stabilization, or linear coupling to oscillatory modes, with a
possible turn-over of the Parker mode, leading in the case of very low
magnetic field to a turn-over with overamplification. The influence of
the physical parameters (strength of differential rotation, relative
strengths of the thermal, magnetic and cosmic rays pressures,
scaleheight of the disk of gas) was pointed out and, as far as possible,
explained.
\par (vi) A transient instability at low azimuthal magnetic field is
found to occur, amplifying both the Alfvenic and Parker modes. A
parallel is drawn with the instability occuring in accretion disks with
a vertical magnetic field.
\par (vii) Realistic parameters, in the case of our galaxy, are shown to
favour a linear coupling between the modes, with no turn-over of the
Parker mode. The sensitivity to the vertical scaleheight of the gas and
magnetic field forbids a categorical conclusion about the possibility of
a turn-over.\\
- One should also discuss the effect of the presence in the galactic
disk  of a strong spiral density wave. Elmegreen (1987) developped a
theory of Parker-Jeans instabilities, mixing the effects of a stratified
field with those of self-gravity. He suggests that the higher density and
lower local shear in the spiral arms might result in strong
instabilities resulting in the formation of large cloud complexes. This
of course reminds of dense structures observed within the arms, in
particular in M51 (Rand \& Kulkarni 1990; Rand 1993).  However,  except
in the vicinity of the corotation radius, the gas should stay in  the
arm only for a fraction of the rotation period, which, with realistic
parameters, is the time scale for the Parker instability. \\
Our results allow us to suggest a different possibility:  we have
described here how a Parker mode emits with high efficiency  (since the
coupling is determined by the shear parameter, which is of the  order of
one) magnetosonic waves, \ie  spiral density waves if we had included
self-gravity, and alfvenic ones.  It can thus be expected (and
preliminary calculations confirm it) that a  spiral wave similarly
generates Parker modes with comparable  efficiency; their growth rate
would be relatively weak,  since they would share the low azimuthal
wavenumber of the spiral, but  the latter would provide a continuous and
strong source for the modes.  Furthermore, the identical azimuthal
wavenumber would provide a natural  explanation for the localization of
the Parker modes in the arms. This mechanism appears as a very promising
way for the spiral wave to inject turbulent energy at large scale in the
interstellar medium, thus feeding the cascade observed down to much
smaller scales. It  will be the subject of a forthcoming paper.

\appendix

\section{Appendix: The dispersion equation and polarizations including
uniform rotation and cosmic-rays}
If the cosmic rays are taken into account, the asymptotic dispersion
equation is changed to~:
\begin{eqnarray}
\Delta(\tom^2)\!\!&\equiv &\!\!(1+2\alpha)\tom^4
\!\!-\!\!2\left\{(1+\alpha)\bk_y^2+(1+\alpha+\beta)
{\alpha-\beta\over4\alpha}\right\} \tom^2\nonumber\\ &
&\hspace{2cm}+\bk_y^2(\bk_y^2-\bk_{\rm P}^2), \label{delta}
\end{eqnarray}
\begin{equation}
{\rm with~~~}\bk_{\rm P}^2\equiv
(1+\alpha+\beta){\alpha+\beta\over2\alpha}>{1\over2}.
\end{equation}
The general dispersion equation is written
$Q_3(\tom^2)=\bk_x^2\Delta(\tom^2)$, with
\begin{eqnarray}
Q_3(\tom^2)&\equiv &(\tom^2-\bk_y^2-\tilde\Omega^2)Q_2(\tom^2)
-k_y^2\tilde\Omega^2(\tom^2+\bk_{\rm Q}^2-\bk_y^2)\label{eq_3b}\\
Q_2(\tom^2)&\equiv &2\alpha
\tom^4-(1+2\alpha)\left(\bk_y^2+{1\over4}\right)\tom^2+
\bk_y^2(\bk_y^2-\bk_{\rm Q}^2).\label{eq_2}
\end{eqnarray}
Here, $\bk_{\rm Q}^2\equiv \bk_{\rm P}^2-1/4$ appears as the maximum
azimuthal wavenumber allowing the Parker instability at $k_x=0$.\\
According to the asymptotic dispersion equation, the maximum growth
rate and the optimum azimuthal wavenumber are~:
\begin{eqnarray}
|\tom_{\Delta_m}|(\alpha,\beta)&=&{\alpha+\beta\over 2\alpha}{ 1\over
1+(1+\alpha+\beta)^{-1/2}}\in]{1\over4},{1\over2}+ {\beta\over2\alpha}[
\label{omopt}\\
\bk_\Delta(\alpha,\beta)&=&\left({\bk_{\rm P}^2\over2}-
(1+\alpha)|\tom_{\Delta_m}|^2 \right)^{1/2}<\bk_{\rm Q}\label{kyopt}
\end{eqnarray}
The dimensionless function outlining the dissymmetrical effect of
rotation is given by~:
\begin{equation}
f_{o}(\alpha,\beta,\bk_y)=(1+2\alpha+2\beta) {\bk_y|Y|\over
\bk_y^2+(1+2\beta)Y^2},\nonumber
\end{equation}
where $Y^2<0$ satisfies $Q_2(Y^2)=0.$  $f_{o}$ is infinite if
$k_y=k_{cr}$ and vanishes at $k_y=k_{\rm Q}$. Although the radial
component of the Coriolis force becomes small compared to each radial
pressure force when $Hk_x\gg \bk_\infty$, it remains dominant if the sum
of these pressure forces vanishes. This occurs at $k_y=k_{cr}$.\\
The polarization of the eigenvector corresponding to the mode $\tom$
satisfying the dispersion equation $Q_3(\tom^2)=\bk_x^2\Delta(\tom^2)$
is~:
\begin{eqnarray}
{\xi_x\over i\xi_z}&=&{-2Q_2(\tom)\over
i\tom\tilde\Omega\bk_y(1+2\alpha+2\beta)-\bk_x(\bk_y^2+(1+2\beta)\tom^2)}
\nonumber\\
{\xi_x\over \xi_y}&=&{-Q_2(\tom)\over 2\alpha
i\tom\tilde\Omega\left(\tom^2-\bk_y^2-{1+2\alpha\over8\alpha}
\right)-\bk_x\bk_y\left(\tom^2-\bk_y^2+
{\beta(1+\alpha+\beta)\over2\alpha}\right)}\nonumber
\end{eqnarray}
In particular, we can check with these formula that the asymptotic
magnetosonic mode is essentially polarized along ${\bf x}$ when
$|\bk_x|\to +\infty$ ($\omega^2\sim (a^2+V_{\rm A}^2)k_x^2$), and that
the two asymptotic Alfvenic modes ($\Delta(\tom^2)=0$) are essentially
polarized in the ($y,z$)-plane.\\

\section{Appendix: Laplace Transform of the solution with shear}
We define the dimensionless epicyclic frequency as~:
\begin{equation}
\tilde\kappa^2={\kappa^2H^2\over V_{\rm A}^2}=\tilde\Omega^2\left(
1+{A\over\Omega}\right)=\tilde\Omega^2-\bk_A^2.\nonumber
\end{equation}
In order to reduce the degree of the differential system
(\ref{systdiff3}), we introduce the convenient function~:
\begin{equation}
\tx_w(k_x,t)=-\left(1+{1\over2\alpha}\right)\bar
K_x\tx_x-{\bk_y\over2\alpha}\tx_y- i{1+2\beta\over
4\alpha}\tx_z.\nonumber
\end{equation}
The only time-dependent parameter, $K_x$, now multiplies the functions
$\tx_x$ and $\tx_w$ only. This leads us to write two equations without
any $K_x$-multiplication, and two other equations expressing $K_x\tx_x$
and $K_x\tx_w$ in terms of $\tx_w,\tx_x,\tx_y,\tx_z$. These latter are
written~:
\begin{eqnarray}
HK_x \tx_w &=& {H^2\over V_{\rm A}^2}\left({\p^2\over\p
t^2}+4A\Omega+k_y^2V_{\rm A}^2\right)\tx_x-2{\Omega H^2\over V_{\rm
A}^2} {\p\over\p t}\tx_y\label{Kwx1},\\
HK_x\tx_x &=& -{1\over 1+2\alpha}\left(2\alpha\tx_w+ Hk_y\tx_y +
i\left({1\over2}+\beta\right)\tx_z\right). \label{Kwx2}
\end{eqnarray}
Performing at this point the Laplace transform (\ref{Laplace}) changes
the $K_x$-multiplications into $\omega$-derivatives, and
time-derivatives change to simple $\omega$-multiplications, with some
additional $\omega$-polynoms whose coefficients depend linearly on the
initial conditions $\tx(k_x,t=0)$ and ${\p\over\p t}\tx(k_x,t=0)$.\\
The first couple of equations, independent of $K_x$,  relate $\bx_y$ and
$\bx_z$  to ($\bx_w,\bx_x$) in a simple algebraic manner~:
\begin{equation}
\Delta \left|
\begin{array}{l}
\bx_y\\
\bx_z
\end{array}
\right. = \left(
\begin{array}{cc}
q_a & q_b\\
q_c & q_d
\end{array}
\right)
\left|
\begin{array}{l}
\bx_w\\
\bx_x
\end{array}
\right.
+\e^{i{\tom\tilde\Omega\bk_x\over\bk_y\bk_A^2}} \left|
\begin{array}{l}
Q_y\\ Q_z
\end{array}
\right..\label{qabcd}
\end{equation}
$Q_y$ and $Q_z$ are polynoms of $\tom$ whose coefficients depend
linearly on the initial conditions. The polynoms ($q_a,q_b,q_c,q_d$) are
independent of $\bk_x$, and constitute a matrix which is invertible as
long as $\Delta\not=0$ and $\tom\not=0$~:
\begin{eqnarray}
q_a&=&-\bk_y\left(\tom^2-\bk_y^2
+(1+\alpha+\beta){\beta\over2\alpha}\right),\nonumber\\
q_b&=&i\tilde\Omega \tom\left((1+2\alpha)(\bk_y^2-\tom^2)+
(1+\alpha+\beta){\alpha-\beta\over 2\alpha}\right),\nonumber\\
q_c&=&{i\over2}(\bk_y^2+(1+2\beta)\tom^2),\nonumber\\
q_d&=&-\tilde\Omega \tom\bk_y(1+\alpha+\beta).\nonumber
\end{eqnarray}
Equations (\ref{Kwx1})-(\ref{Kwx2}) may be expressed as a differential
system of order two on the functions  ($\bx_w,\bx_x$), using
Eq.~(\ref{qabcd}).
\begin{equation}
{\bk_A^2\bk_y\over\tilde\Omega}\Delta{\p\over\p \tom}
\left|
\begin{array}{l}
\bx_w\\
\bx_x
\end{array}
\right. = \left(
\begin{array}{cc}
p_a & p_b\\
p_c & p_d
\end{array}
\right) \left|
\begin{array}{l}
\bx_w\\
\bx_x
\end{array}
\right. +\e^{i{\tom\tilde\Omega\bk_x\over\bk_y\bk_A^2}} \left|
\begin{array}{l}
P_w\\
P_x
\end{array}
\right., \label{diffabcd}
\end{equation}
where ($p_a,p_b,p_c,p_d$) are simple polynoms of $\tom$ whose
coefficients are independent of $\bk_x$.\\
$P_w$ and $P_x$ are two polynoms of $\tom$ of respective degree 5 and 3,
whose coefficients depend linearly on the six initial conditions
\begin{equation}
\left(\tx_x,\tx_y,\tx_z,{\p\over\p t}\tx_x,{\p\over\p t}\tx_y, {\p\over\p
t}\tx_z\right)(k_x,t=0),\nonumber
\end{equation}
independently from the $\bk_x$ and $\bk_A$ parameters.\\
The coefficients of the matrix in Eq.~(\ref{diffabcd}) are given by~:
\begin{eqnarray}
p_a&=&\tilde\Omega\bk_y \tom\left(\tom^2-\bk_y^2 +(1+\alpha+\beta)
{\beta\over2\alpha}\right)\nonumber\\
p_b&=&-i\{(\tom^2-\bk_y^2 -\tilde\kappa^2)\Delta(\tom^2)
-\tilde\Omega^2k_y^2(\tom^2-\bk_y^2+\bk_{\rm P}^2)\}\nonumber\\
p_c&=&-iQ_2(\tom^2)\nonumber\\
p_d&=&-p_a\nonumber
\end{eqnarray}
In particular, the determinant of (\ref{diffabcd}) is
$p_ap_d-p_bp_c=\Delta Q_3$, where the polynom $Q_3$ defined in
(\ref{eq_3b}) is modified by the differential force as~:
\begin{eqnarray}
Q_3(\tom^2)=(\tom^2-\bk_y^2-\tilde\kappa^2)
Q_2(\tom^2)-k_y^2\tilde\Omega^2(\tom^2+\bk_{\rm Q}^2-\bk_y^2).
\label{dispshear}
\end{eqnarray}
$\Delta$ and $Q_2$ are respectively defined in (\ref{delta}) and
(\ref{eq_2}).\\
The differential system (\ref{diffabcd}) can easily be transformed, by
substitution, into the differential equation of second order
(\ref{Frobenius}).\\
We derive for $\bx_x$~:
\begin{eqnarray}
f_x(\tom^2)&=&{\Delta\over p_c}\label{fx}\\
g_x(\tom^2)&=&{\tilde\Omega^2\over\bk_A^4\bk_y^2}\left\{{Q_3\over\Delta}
-{\bk_A^2\bk_y\over\tilde\Omega}{p_c\over\Delta}
{\p\over\p\tom}{p_d\over p_c}\right\}\label{gx}\\
h_x(\bk_x,\tom)&=&{\tilde\Omega^2\over\bk_A^4\bk_y^2}
\e^{i{\tom\tilde\Omega\bk_x\over\bk_y\bk_A^2}}
\left\{{P_wp_c-P_x(p_a-i\bk_x\Delta)\over \Delta^2}\right.\nonumber\\
& &\hspace{3cm}+ \left.{\bk_A^2\bk_y\over\tilde\Omega}{p_c\over\Delta}
{\p\over\p\tom}{P_x\over p_c}\right\}\label{hx}
\end{eqnarray}
In particular, the polynom $\Delta$ factors ($P_wp_c-P_xp_a$), which
ensure that $f_xh_x$ is regular at $\tod$. The functions ($f_w,g_w,h_w$)
are defined as ($f_x,g_x,h_x$), replacing $p_a$ with $p_d$, $p_b$ with
$p_c$, $P_w$ with $P_x$ and conversely.\\
Inverting the system (\ref{qabcd}) allows us to transform the
differential system (\ref{diffabcd}) into a similar one verified by
($\bx_y,\bx_z$)~:
\begin{equation}
{\bk_A^2\bk_y\over\tilde\Omega}\tom\Delta{\p\over\p
\tom} \left|
\begin{array}{l}
\bx_y\\
\bx_z
\end{array}
\right.  = \left(
\begin{array}{cc}
p_a' & p_b'\\
p_c'& p_d'
\end{array}
\right) \left|
\begin{array}{l}
\bx_y\\
\bx_z
\end{array}
\right. +\e^{i{\tom\tilde\Omega\bk_x\over\bk_y\bk_A^2}} \left|
\begin{array}{l}
P_y\\
P_z
\end{array}
\right.. \label{difyzabcd}
\end{equation}
The four polynoms of degree 3 in $\tom^2$, ($p_a',p_b',p_c',p_d'$) are
independent of $k_x$. $P_y$ and $P_z$ are polynoms of degree 5 in
$\tom$, whose coefficients depend linearly on the initial conditions.\\
The major properties of the matrix (\ref{difyzabcd}) are~:
\par (i) $\tom^2\Delta(\tom^2)$ factors its determinant~:
\begin{equation}
p_a'p_d'-p_b'p_c'=\tom^2\Delta Q_3'.\nonumber
\end{equation}
\par (ii)  Its trace is given by~:
\begin{equation}
p_a'+p_d'=-{\bk_A^2\bk_y\over\tilde\Omega}\tom\Delta\left(
{\p\log\Delta\over\p\tom}-{1\over\tom}\right).\nonumber
\end{equation}
The polynom $Q_3'(\tom^2)$ of degree 3 in $\tom^2$ identifies with
$Q_3(\tom^2)$ when the parameter of shear $\bk_A$ vanishes, so that the
determinant of the matrix (\ref{difyzabcd}) gives the classical
dispersion equation ($Q_3(\tom^2)-\bk_x^2\Delta(\tom^2)=0$) of
Sect.~5.\\
The second order differential equations satisfied by $\bx_y$
and $\bx_z$, derived from (\ref{difyzabcd}), are somewhat different~:
\begin{eqnarray}
f_z(\tom^2)&=&{\Delta^2\over p_c'}\label{fz}\\
g_z(\tom^2)&=&{\tilde\Omega^2\over\bk_A^4\bk_y^2}
\left\{{Q_3'\over\Delta}-{2\bk_A^2\bk_y\over\tilde\Omega}
{p_c'\over\Delta} {\p\over\p\tom^2}{p_d'\over p_c'}\right\}\label{gz} \\
h_z(\bk_x,\tom)&=&{\tilde\Omega^2\over\bk_A^4\bk_y^2}
\e^{i{\tom\tilde\Omega\bk_x\over\bk_y\bk_A^2}}\left\{{P_yp_c'
-P_z(p_a'-i\bk_x\tom\Delta)\over\tom^2\Delta^2}\right.\nonumber\\
& &\hspace{3cm}+\left.{\bk_A^2\bk_y\over\tilde\Omega}
{p_c'\over\tom\Delta} {\p\over\p\tom}{P_z\over p_c'}\right\}\label{hz}
\end{eqnarray}
(replace $p_d'$ with $p_a'$, $p_c'$ with $p_b'$, $P_y$ with $P_z$ and
conversely to get  $f_y,g_y$and $h_y$.)\\
Let us define $\bx_{\rm H}(\tom)$, solution of the associated
homogeneous differential equation, which is $k_x$-independent~:
\begin{equation}
\left\{{\p^2\over\p\tom^2} + {\p\log f\over\p\tom}{\p\over\p\tom}+
g\right\}\bx_{\rm H}=0. \label{homogenous}
\end{equation}
Using the WKB approximation (\ref{WKB}) when $\tom\to\infty$, the
asymptotic behaviour of $\bx_{\rm H}$ is~:
\begin{equation}
\bx_{\rm H}(\tom)\sim {1\over\tom^n}\exp\left\{\pm i
{\tilde\Omega\over\bk_A^2\bk_y}\left({2\alpha\over1+
2\alpha}\right)^{1/2} {\tom^2\over2}\right\} ,\label{ginfini}
\end{equation}
with $n=-1/2$ for $\bx_{hw}$, $n=1/2$ for $\bx_{hx}$, and  $n=3/2$ for
$\bx_{hy}$ and $\bx_{hz}$.\\
Let us define $\bx_\ominus(\tom)$ and  $\bx_\oplus(\tom)$ as the
solutions which respectively converge at ($\tom\to ip-\infty$) and
($\tom\to ip+\infty$).\\
Referring to the typical case of parabolic cylindrical differential
equation, we assume that a homogeneous solution which vanishes at both
($\tom\to ip-\infty$) and ($\tom\to ip+\infty$) may exist only for a
discrete set of the parameters
($\bk_y,\alpha,\beta,\tilde\Omega,\bk_A$). In the general case,
($\bx_\ominus,\bx_\oplus$) constitute a basis of solutions, and a
complex constant $c$ exists, independent from $k_x$, so that
\begin{equation}
\bx_\ominus {\p\bx_\oplus\over\p\tom}-
{\p\bx_\ominus\over\p\tom}\bx_\oplus={1\over c f(\tom)}.\nonumber
\end{equation}
Defining the function containing the information about the initial
radial shape of the perturbation $L_c(\bk_x,\tom)$ as
\begin{equation}
L_c(\bk_x,\tom)= cf(\tom)h(\bk_x,\tom),\label{Lc}
\end{equation}
the general solution of the differential equation (\ref{Frobenius})
is
\begin{eqnarray}
\bx(k_x,\tom)=\bx_\oplus(\tom)
 \int^{\tom} L_c(\bk_x,\tom')\bx_\ominus(\tom') \d\tom' \nonumber\\
-\bx_\ominus(\tom) \int^{\tom}L_c(\bk_x,\tom')\bx_\oplus(\tom')
\d\tom',
\end{eqnarray}
The integration limits must be chosen so as to allow an inverse Laplace
transform (\ref{invLaplace}). The asymptotic behaviour of
($\bx_\ominus,\bx_\oplus$) leads to the unique possible choice,
\begin{eqnarray}
\bx(k_x,\tom)=\bx_\oplus(\tom)
 \int_{ip-\infty}^{\tom} L_c(\bk_x,\tom')\bx_\ominus(\tom') \d\tom'
\nonumber\\
+\bx_\ominus(\tom)
\int_{\tom}^{ip+\infty}L_c(\bk_x,\tom')\bx_\oplus(\tom') \d\tom'
,\label{general}
\end{eqnarray}
As a proof, first notice that when $Re(\tom)\to\pm\infty$
\begin{equation}
\bx_\oplus(\tom)\int_{ip-\infty}^{\tom}L_c(\bk_x,\tom')
\bx_\ominus(\tom') \d\tom'=O\left({\bx_\oplus \bx_\ominus
L_c\over\tom}\right).\nonumber
\end{equation}
Using the same index $n$ as in (\ref{ginfini}), we obtain
\begin{equation}
\bx(k_x,\tom)=O\left({L_c\over\tom^{2n+1}}\right).\nonumber
\end{equation}
Considering the degree of the polynoms involved in Eqs.~(\ref{hx}) and
(\ref{hz}), we get the crude estimation $L_c(\bk_x,\tom)=O(\tom^{2n})$.
We conclude that when  $Re(\tom)\to\pm\infty$,
\begin{equation}
\bx(k_x,\tom)=O\left({1\over\tom}\right), \nonumber
\end{equation}
which ensures the convergence of the inverse Laplace transform
(\ref{invLaplace}).\\

\section{Appendix: $\tom$-singularities and time-asymptotic behaviour}
We can read on the differential system (\ref{diffabcd}) that the
functions ($\bx_w,\bx_x$) are regular everywhere except where
$\Delta(\tom)=0$. Let us note $\tom_j$ one of the four solutions of this
equation ($\tom_j\in\{\pm\tod,\pm\todp\}$).\\
We deduce from equation (\ref{qabcd}) that ($\bx_y,\bx_z$) have the same
field of regularity as  ($\bx_w,\bx_x$). Hence, the various other
singularities of $f$ and $g$ appearing in (\ref{fx})-(\ref{gx}) and
(\ref{fz})-(\ref{gz}), as well as the apparent singularity at $\tom=0$
of the differential system (\ref{difyzabcd}), are not true singularities
of the solution $\bx$.\\
To analyse the singularity of a differential equation, Frobenius method
requires an expansion of the functions $f$ and $g$ near the singular
point $\tom_j$.
\begin{eqnarray}
f(\tom)&=&{f_{-1}^j\over\tom-\tom_j}+f_{0}^j+
f_1^j(\tom-\tom_j)+O(\tom-\tom_j)^2\nonumber\\
g(\tom)&=&{g_{-2}^j\over(\tom-\tom_j)^2}+{g_{-1}^j\over\tom-\tom_j}
+g_{0}^j+O(\tom-\tom_j)\nonumber
\end{eqnarray}
The indicial exponent $\alpha_j$ is defined as~:
\begin{equation}
\alpha_j^2+(f_{-1}-1)\alpha_j+g_{-2}=0.\nonumber
\end{equation}
We deduce from Eqs.~(\ref{fx})-(\ref{gx}) and (\ref{fz})-(\ref{gz}) that
\begin{eqnarray}
(f_{-1w}^j,g_{-2w}^j)=(f_{-1x}^j,g_{-2x}^j)&=&(1,0), \nonumber\\
(f_{-1y}^j,g_{-2y}^j)=(f_{-1z}^j,g_{-2z}^j)&=&(2,0). \nonumber
\end{eqnarray}
Consequently, the indicial exponents are (0,0) for $\bx_{hw}$ and
$\bx_{hx}$, and (-1,0) for $\bx_{hy}$ and $\bx_{hz}$.\\
A straightforward consequence is the existence of a homogeneous
solution, $\bx_0^j$ which is regular at $\tom_j$, with
$\bx_0^j(\tom_j)=1$.\\
Frobenius method leads us to define another regular function $\bx_1'^j$
and two complex constants $\gamma_y^j,\gamma_z^j$, so that a second
homogeneous solution $\bx_1^j$ is written~:
\begin{eqnarray}
\bx_{1w}^j(\tom)&=&\bx_{0w}^j(\tom)\ln(\tom-\tom_j)+ \bx_{w1}'(\tom)
\label{xiw1}\\ \bx_{1x}^j(\tom)&=&\bx_{0x}^j(\tom)\ln(\tom-\tom_j)+
\bx_{x1}'(\tom) \label{xix1}\\
\bx_{1y}^j(\tom)&=&{1 \over
\tom-\tom_j}+\gamma_y^j\bx_{0y}^j(\tom)\ln(\tom-\tom_j)+
\bx_{y1}'^j(\tom) \label{xiy1}\\ \bx_{1z}^j(\tom)&=&{1 \over
\tom-\tom_j}+\gamma_z^j\bx_{0z}(\tom)^j\ln(\tom-\tom_j)+
\bx_{z1}'^j(\tom) \label{xiz1}
\end{eqnarray}
$\bx_0^j$ and $\bx_1^j$ constitute a local basis of homogeneous
solutions. A complex constant $c^j$ consequently exists,  defined as~:
\begin{equation}
\bx_1^j{\p\bx_0^j\over\p\tom}-\bx_0^j{\p\bx_1^j\over\p\tom} ={1\over
c^jf(\tom)},\nonumber
\end{equation}
A local expansion of this equation near $\tom_j$ leads to
\begin{eqnarray}
c^j_x&=&-p_c(\tom_j)\left[{\p\Delta\over\p\tom}(\tom_j)\right]^{-1}
\label{cx}\\
c^j_y&=&p_b'(\tom_j)\left[{\p\Delta\over\p\tom}(\tom_j)\right]^{-2}
\nonumber\\
c^j_z&=&p_c'(\tom_j)\left[{\p\Delta\over\p\tom}(\tom_j)\right]^{-2}
\nonumber
\end{eqnarray}
The general solution of the differential equation (\ref{Frobenius}) can
be written
\begin{eqnarray}
\bx(k_x,\tom)=\bx_0^j(\tom) \int_{l_1^j}^{\tom}
L_{c^j}(\bk_x,\tom')\bx_1^j(\tom')\d\tom' \nonumber\\
-\bx_1^j(\tom)\int_{l_0^j}^{\tom} L_{c^j}(\bk_x,\tom')\bx_0^j(\tom')
\d\tom'.\label{local}
\end{eqnarray}
with appropriate limits of integration $l_1^j$ and $l_0^j$, and
\begin{equation}
L_c^j(\bk_x,\tom)= c^jf(\tom)h(\bk_x,\tom).
\end{equation}
Equations (\ref{fx})-(\ref{hx}) and (\ref{fz})-(\ref{hz}) show that
$L_{c^j_x},L_{c^j_y}$ and $L_{c^j_z}$ are regular at $\tom_j$. A local
study of (\ref{local}) leads to the local expansions mentioned in
Sect.~8.2.1 with~:
\begin{eqnarray}
\Lambda^j(k_x)&=&-\int_{l_{0}^j}^{\tom_j}
L_{c^j}(\bk_x,\tom')\bx_{0}^j(\tom')\d\tom', \nonumber\\
\Gamma^j(k_x)&=&L_c^j(\tom_j)+\gamma^j\Lambda^j(k_x)\nonumber.
\end{eqnarray}
We can get rid of the undefined integration limit $l_{0}^j$, using the
converging homogeneous solutions $\bx_\ominus,\bx_\oplus$ defined in
Appendix~B. Let us project them as~:
\begin{eqnarray}
\bx_\ominus&=&\lambda_-\bx_0+\mu_-\bx_1\nonumber\\
\bx_\oplus&=&\lambda_+\bx_0+\mu_+\bx_1\nonumber
\end{eqnarray}
Let us, for the sake of clarity, forget the index ($j,x$), and
concentrate from now on on the first component $\bx_x$ near the
``unstable'' singularity $\tod$.\\
We deduce from Eq.~(\ref{general})
\begin{eqnarray}
\Lambda^j_x(k_x)&=&{\mu_+\over
\mu_-\lambda_+-\mu_+\lambda_-}\int_{ip-\infty}^{\tom_1}
L_{c^j}\bx_\ominus-\int_{\tom_1}^{\tod}L_{c^j}\bx_0\nonumber\\
&+&{\mu_-\over \mu_-\lambda_+-\mu_+\lambda_-}\int_{\tom_1}^{ip+\infty}
L_{c^j}\bx_\oplus. \nonumber
\end{eqnarray}
We use an integration by parts~:
\begin{eqnarray}
\int L_c(\bk_x,\tom')\bx(\tom')\d\tom' ={c\tilde\Omega\over\bk_A^2\bk_y}
\left[\e^{i{\tom\tilde\Omega\bk_x\over\bk_y\bk_A^2}}{P_x\over
p_c}\right]+\nonumber\\
{c\tilde\Omega^2\over\bk_A^4\bk_y^2}\int
\e^{i{\tom\tilde\Omega\bk_x\over\bk_y\bk_A^2}}\left\{\left({P_wp_c-P_xp_a\over\Delta
p_c}\right)\bx-{\bk_A^2\bk_y\over\tilde\Omega}{P_x\over
p_c}{\p\bx\over\p\tom} \right\} \d\tom' \label{bypart}
\end{eqnarray}
In particular, we know from Appendix~B that the singularity at
$p_c(\tom)=0$ is artificial, and that $\Delta(\tom^2)$ always divides
$(P_wp_c-P_xp_a)(k_x,\tom)$.\\
In order to calculate ${\cal C}_{\rm PP}$, we normalize the initial
perturbation contained in  $P_x,P_w$ so that
\begin{equation}
{\cal C}_{\rm PP} =\lim_{k_{0x}\to-\infty}{2\bk_y AH\over V_A}
e^{-i{\tom\tilde\Omega\bk_{0x}\over\bk_y\bk_A^2}} \Lambda^{\rm
P}_x(k_{0x})
\end{equation}
The only non-vanishing contributions, when $k_{0x}\to-\infty$, come from
the pole at $\tod$ appearing in $\p\bx_1/\p\tom$, and the integrated
term in the brackets in Eq.~(\ref{bypart}). Using (\ref{cx}), we finally
obtain~:
\begin{equation}
{\cal C}_{\rm PP}= P_x(\tod)
\left[{\p\Delta\over\p\tom}(\tod)\right]^{-1}\left(1+
{2i\pi\mu_-\mu_+\over \mu_-\lambda_+-\mu_+\lambda_-}\right).
\label{analytic}
\end{equation}
with
\begin{eqnarray}
P_x(\tod)&=&i\tod\left\{2\bk_y\left[\tod^2-\bk_y^2+(1+\alpha+\beta)
{\beta\over2\alpha}\right] \xi_{{\rm P}_y}(\infty)\right.\nonumber\\
&+&\left.\left[(2\beta+1)\tod^2+\bk_y^2\right] i\xi_{{\rm
P}_z}(\infty)\right\}. \nonumber
\end{eqnarray}
where $[0,\xi_{{\rm P}_y}(\infty),\xi_{{\rm P}_z}(\infty)]$ is the
limit, when $k_x\to\pm\infty$ of the Parker polarization $\xi_{\rm
P}(k_x)=(\xi_{{\rm P}_x},\xi_{{\rm P}_y},\xi_{{\rm P}_z})$, given in
Sect.~5, and normalized with $\xi_{{\rm P}_x}=1/\bk_x$. Hence
$P_x(\tod)$ does not depend on the strength of differential rotation.\\
While $\bx_0$ was defined in a unique way ($\bx_0(\tod)=1$), the second
homogeneous solution $\bx_1$ (defined by its leading singular behaviour
at $\tod$) may be arbitrarily replaced with $\bx_1+\lambda'\bx_0$. We
can easily check that the formula (\ref{analytic}) does not depend on
such a normalization.\\
Assuming that $\bx_\ominus(\tom)$ diverges when Re$(\tom)\to+\infty$, we
deduce from the parity of the homogeneous differential
Eq.~(\ref{homogenous}) that
$\bx_\oplus(\tom)\equiv\bx_\ominus^*(-\tom^*)$ is independent from
$\bx_\ominus$ and converges when Re$(\tom)\to+\infty$. This allows us to
choose
\begin{eqnarray}
\mu_+&=&\mu_-^*,\nonumber\\
\lambda_+&=&\lambda_-^*,\nonumber
\end{eqnarray}
and obtain Eq.~(\ref{analyticturn}).\\

\section{Appendix: WKB validity}
$\hat\xi_\pm$ are the {\it exact} solutions of the differential equation
\begin{eqnarray}
\left\{{\p^2\over\p\tom^2} + {\p\log f\over\p\tom}{\p\over\p\tom}+
g+{\p^2\log W\over4\;\p\tom^2}-\left( {\p\log W\over4\;\p\tom}
\right)^2\right\}\hat\xi_\pm=0.\nonumber
\end{eqnarray}
The quality of the approximation is consequently measured by the
following criterion~:
\begin{equation}
\left|{1\over4}{\p^2\log W\over\p\tom^2}-\left({1\over4}{\p\log
W\over\p\tom} \right)^2\right|\ll g.\label{criterWKB}
\end{equation}
A straightforward result is the estimation of $\xi_H$ when
$|\tom|\to\infty$.\\
Equations (\ref{fx})-(\ref{gx}) and (\ref{fz})-(\ref{gz}) give the
asymptotic behaviour of $f(\tom^2)$ and $g(\tom^2)$. We deduce, when
$|\tom|\to\infty$~:
\begin{equation}
W(\tom^2)\sim g(\tom^2) \sim {\tilde\Omega^2\over\bk_A^4\bk_y^2}
{2\alpha\over1+2\alpha}\tom^2+O(1).\nonumber
\end{equation}
The criterion (\ref{criterWKB}) simplifies into
\begin{equation}
\tom\gg\left({3\bk_A^4\over8\tilde\Omega^2\bk_y^2}
{1+\alpha\over\alpha}\right)^{1/4},\nonumber
\end{equation}
and hence justifies the estimation (\ref{ginfini}) used in Appendix~B~:
\begin{equation}
\hat\xi_H(\tom)\sim {1\over(\tom f(\tom^2))^{1/2}}\exp\left\{\pm i
{\tilde\Omega\over\bk_A^2\bk_y}\left({2\alpha\over1+
2\alpha}\right)^{1/2} {\tom^2\over2}\right\} .
\end{equation}
The WKB approximation fails to describe the neighbourhood of the
following points in the complex plane ~:
\par (i) the points where $W(\tom^2)=0$ are singularities, artificially
introduced by the WKB formalism.
\par (ii) the points where $f$ or $g$ are singular lack interest,
because they are known to be artificial (Appendix~D).
\par  (iii) the real singularity of $\hat\xi$, $\tod$, is better
described by the local analysis performed in Sect.~8.2. and
Appendix~E.\\
Apart from these singular points, the WKB-criterion (\ref{criterWKB})
may be satisfied for bounded values of $\tom$ if
\begin{eqnarray}
{\bk_A^2\over\tilde\Omega}\ll1.
\end{eqnarray}
Indeed, this condition implies that far from the singular points,
$W(\tom^2)\sim g(\tom^2),$ and
\begin{equation}
{1\over g^{3/2}}{\p g\over\p\tom}\ll1\mbox{ and }  {1\over g^2}{\p^2
g\over\p\tom^2}\ll1,\nonumber
\end{equation}
which leads to (\ref{criterWKB}). Before detailing the physical meaning
of the WKB-criterion, let us recall the simplifications allowed by the
estimation of the function $\tx(k_x,t)$.\\
We project the homogeneous solutions ($\hat\xi_\ominus,\hat\xi_\oplus$)
onto the WKB-basis ($\hat\xi_+,\hat\xi_-$)~:
\begin{eqnarray}
\hat\xi_\ominus(\tom)&=&c_\ominus(\tom)\hat\xi_-(\tom)+d_\ominus(\tom)
\hat\xi_+(\tom),\nonumber\\
\hat\xi_\oplus(\tom)&=&c_\oplus(\tom)\hat\xi_-(\tom)+d_\oplus(\tom)
\hat\xi_+(\tom).\nonumber
\end{eqnarray}
The osculatory coefficients $(c_\ominus,d_\ominus,c_\oplus,d_\oplus)$
are slowly varying functions of $\tom$ where the WKB-approximation is
valid. Their value changes at the crossing of the Stokes lines linked to
the presence of the $W(\tom^2)=0$ singular points in the complex
plane.\\
The convergence properties of ($\hat\xi_\ominus,\hat\xi_\oplus$) are
equivalent to
\begin{equation}
d_\ominus(\tom\to ip-\infty)=c_\oplus(\tom\to ip+\infty)=0.\nonumber
\end{equation}
Replacing ($\hat\xi_\ominus,\hat\xi_\oplus$) with their WKB-expressions
in (\ref{general}) leads us to write the inverse Laplace transform
(\ref{invLaplace}) as~:
\begin{eqnarray}
\tx(k_x,t)={1\over2\pi}\int_{ip-\infty}^{ip+\infty}
\left(\mu_+\e^{i\Psi_+}- \mu_-\e^{i\Psi_-}\right)\d \omega,  \nonumber
\end{eqnarray}
The phase $\Psi_\pm(\tom,t)$ is written in a dimensionless form
\begin{equation}
\Psi_\pm(\tom,t)=-{\tilde\Omega\over\bk_A^2\bk_y}(\bk_x-2\bk_yAt)
\pm\int^{\tom} W^{1/2}(\tom')\d\tom',\nonumber
\end{equation}
and we have defined $(\mu_+,\mu_-)$ as~:
\begin{eqnarray}
\mu_+(\tom,k_x)&=&
{1\over(f^2W)^{1/4}}\left\{c_\oplus\int_{ip-\infty}^{\tom}
L_c\hat\xi_\ominus+c_\ominus\int_{\tom}^{ip+\infty}
L_c\hat\xi_\oplus\right\},\nonumber\\
\mu_-(\tom,k_x)&=& {1\over
(f^2W)^{1/4}}\left\{d_\oplus\int_{ip-\infty}^{\tom}
L_c\hat\xi_\ominus+d_\ominus\int_{\tom}^{ip+\infty}
L_c\hat\xi_\oplus\right\}.\nonumber
\end{eqnarray}
These two latter functions slowly vary with regard to the fast
oscillations of the phase $\exp\{i\Psi_\pm(\tom,t)\}$ in the
WKB-limit.\\
Equations (\ref{fx})-(\ref{gx}) show that the singular points $\tom_s$
associated to the $\hat\xi_x$-differential equation verify
$Q_2(\tom_s)=0$. They are independent of the WKB parameter
$\bk_A^2/\tilde\Omega$ and are well isolated from the root we are
interested in ($\tom_3\sim O(1/\tilde\Omega)\ll1$).\\
On the contrary, the $\hat\xi_w$,$\hat\xi_y$ and
$\hat\xi_z$-differential equations each introduce a singularity
$\tom_s'\sim O(1/\tilde\Omega)\ll1$ comparable with the modified Parker
root $\tom_3$, precluding a direct WKB approximation.\\
Nevertheless, we can recover an approximation of the behaviour of
$\hat\xi_w$,$\hat\xi_y$ and $\hat\xi_z$ from the WKB approximation of
$\hat\xi_x$, using Eqs.~(\ref{diffabcd}) and (\ref{qabcd}).\\


\begin{thebibliography}{}
\bibitem{} Balbus S.A., Hawley J.F. 1991, ApJ 376, 214

\bibitem{} Balbus S.A., Hawley J.F. 1992, ApJ 400, 610

\bibitem{} Beck R. 1991, Proc. IAU Symp. 144, The interstellar
disk-halo connection in galaxies, p. 267

\bibitem{} Blitz L., Shu F.H. 1980, ApJ 238, 148

\bibitem{} Bloemen J.B.G.M. 1987, ApJ 322, 694

\bibitem{} Boulares A., Cox D.P. 1990, ApJ 365, 544

\bibitem{} Cesarsky C.J. 1980, ARA\&A 18, 289

\bibitem{} Chandrasekhar S. 1960, Proc. Nat. Acad. Sci. 46, 253

\bibitem{} Dubrule B., Knobloch E. 1993, A\&A 274, 667

\bibitem{} Elmegreen B.G. 1982, ApJ 253, 634

\bibitem{} Elmegreen B.G. 1982, ApJ 253, 655

\bibitem{} Elmegreen B.G. 1987, ApJ 312, 626

\bibitem{} Elmegreen B.G. 1989, ApJ 342, L67

\bibitem{} Elmegreen B.G. 1991, ApJ 378, 139

\bibitem{} Giz A.T., Shu F.H. 1993, ApJ 404, 185

\bibitem{} Goldreich P., Lynden-Bell D. 1965, MNRAS 130, 125

\bibitem{} Gomez de Castro A.I., Pudritz R.E. 1992, ApJ 395, 501

\bibitem{} Hanawa T., Nakamura F., Nakano T. 1992, PASJ 44, 509

\bibitem{} Horiuchi T., Horiuchi T., Matsumoto R., Shibata K., Hanawa T.
1988, PASJ 40, 147

\bibitem{} Hughes D.W., Cattaneo F. 1987, Geophys. Astrophys. Fluid Dyn.
39, 65

\bibitem{} Lachi\`eze-Rey M., Ass\`eo E., Cesarsky C.J., Pellat R. 1980,
ApJ 238, 175

\bibitem{} Lerche I., Parker E.N. 1967, ApJ 149, 559

\bibitem{} Lin C.C., Thurstans R.P. 1984, Proc. of a Course \& Workshop
on Plasma Astrophysics, Varenna, Italy, ESA SP 207

\bibitem{} Parker E.N. 1966, ApJ 145, 811

\bibitem{} Parker E.N. 1967, ApJ 149, 535

\bibitem{} Parker E.N. 1975, ApJ 201, 74

\bibitem{} Parker E.N. 1992, ApJ 401, 137

\bibitem{} Rand R.J., Kulkarni S.R. 1990, ApJ 349, L43

\bibitem{} Rand R.J. 1993, ApJ 410, 68

\bibitem{} Ruzmaikin A.A., Shukurov A.M., Sokoloff D.D. 1988, ``Magnetic
Fields of Galaxies'', Kluwer Academic Publishers

\bibitem{} Ryu D., Goodman J. 1992, ApJ 388, 438

\bibitem{} Shu F.H. 1974, A\&A 33, 55

\bibitem{} Tagger M., Pellat R., and Coroniti F.V. 1992, ApJ 393, 708

\bibitem{} Tagger M., Pellat R., and Coroniti F.V. 1993, Proc. of the
Fourth International Conference on Plasma Physics and Controlled Nuclear
Fusion, Toki, Japan, ESA SP-351

\bibitem{} Zweibel E.G., Kulsrud R.M. 1975, ApJ 201, 63

\end{thebibliography}
\end{document}